\documentclass[a4paper,fleqn]{cas-dc}

\usepackage[authoryear,longnamesfirst]{natbib}
\usepackage{rotating}
\usepackage{booktabs}

\def\tsc#1{\csdef{#1}{\textsc{\lowercase{#1}}\xspace}}
\tsc{WGM}
\tsc{QE}

\begin{document}
\let\WriteBookmarks\relax
\def\floatpagepagefraction{1}
\def\textpagefraction{.001}

\shorttitle{The State of the Practice in Privacy Impact Assessments}

\shortauthors{}  

\title [mode = title]{The State of the Practice in Privacy Impact Assessments: Understanding the Professionals' Perspectives, Experiences, and Current Procedures}


\tnotemark[1] 

\tnotetext[1]{} 

%

\author[1]{Samuel Wairimu}[orcid=0000-0003-1750-649X]

\cormark[1]


\ead{samuel.wairimu@kau.se}


\credit{Methodology, Validation, Formal analysis, Investigation, Data Curation, Writing - Original Draft, Writing - Review \& Editing, Visualization}

\affiliation[1]{organization={Department of Mathematics and Computer Science, Karlstad University},
            addressline={Universitetsgatan 2}, 
            city={Karlstad},
            postcode={651 88}, 
            state={V\"{a}rmland},
            country={Sweden}}

\author[1]{Leonardo Horn Iwaya}[orcid=0000-0001-9005-0543]


\ead{leonardo.iwaya@kau.se}


\credit{Conceptualization, Methodology, Validation, Formal analysis, Investigation, Data Curation, Writing - Original Draft, Writing - Review \& Editing, Visualization}

\author[2]{Ala Sarah Alaqra}[orcid=0000-0002-6509-3792]


\ead{as.alaqra@kau.se}


\credit{Conceptualization, Methodology, Validation, Formal analysis, Investigation, Writing - Original Draft, Writing - Review \& Editing}

\affiliation[2]{organization={Department of Information Systems, Karlstad University},
            addressline={Universitetsgatan 2}, 
            city={Karlstad},
            postcode={651 88}, 
            state={V\"{a}rmland},
            country={Sweden}}

\cortext[1]{Corresponding author: Samuel Wairimu}



\begin{abstract}
Organisations frequently utilise Privacy Impact Assessments (PIAs), Data Protection Impact Assessments (DPIAs), or even Fundamental Rights Impact Assessments (FRIAs) to identify and evaluate potential risks and possible infringements of fundamental rights. While the adoption of PIAs and DPIAs has grown significantly in recent years, robust scientific evidence regarding their operational challenges, implementation strategies, and tangible benefits remains scarce. In particular, empirical research conducted post-enactment of the General Data Protection Regulation (GDPR) is notably limited -- especially studies that directly interview professionals about their PIA/DPIA practices or their engagement with FRIAs. To do so, in this paper we leveraged semi-structured interviews to elicit privacy professionals' perspectives on PIA/DPIA activities and analysed the responses using reflexive thematic analysis. As a result, this work provides a comprehensive thematic framework for the field, offering a rich and thick description of professionals' views and experiences. In total, 26 participants from 17 countries were interviewed, revealing how they plan and conduct PIAs/DPIAs, including how they assess trade-offs, the benefits and challenges they perceive, and how they view the interplay between PIAs/DPIAs and the EU AI Act. Among the main findings were low adherence to established DPIA methodologies; differences and similarities between PIAs and DPIAs; a perceived neglect of data subjects' interests during DPIA processes; discussions on building FRIAs on top of DPIAs; and the potential for holistic impact assessments, though accompanied by concerns regarding assessment fatigue. Thus, this study informs privacy professionals about real-world challenges and experiences of PIA/DPIA implementation, and guides researchers toward future research directions.
\end{abstract}


\begin{highlights}
\item The EU GDPR has propelled research and practice in PIAs and DPIAs worldwide.
\item Organisations often employ \textit{ad hoc} PIAs and DPIAs, adapted to their processes.
\item The EU AI Act further stresses the need for DPIAs but also introduces new requirements for FRIAs.
\item Trade-offs among protection goals are rarely considered as part of assessments.
\item More guidance from authorities is expected for FRIAs under the EU AI Act.
\end{highlights}




\maketitle

\section{Introduction}\label{intro}
Advancements in technology, including the widespread availability of mobile Internet and the rise of generative AI systems, are enhancing everyday life by improving connectivity, efficiency, and quality across various sectors of society. However, along with these benefits, technology can inadvertently introduce risks, such as potential violations of fundamental rights, particularly regarding privacy, through the processing of personal data. Therefore, it is essential to critically evaluate the impacts of new technologies alongside their development, proactively adopting strategies to identify problems and influence system design. As noted by \cite{suzor2019human}, systems that are not designed to combat prejudice, bias, and discrimination often end up amplifying these issues. This logic can also be applied to harms related to other rights and freedoms.

Among the prominent strategies for managing the risks associated with new technologies, impact assessments have historically been viewed as valuable proactive measures~\citep{clarke2009privacy}. For instance, to address privacy concerns, organisations may conduct Privacy Impact Assessments (PIAs) to identify and mitigate risks. PIAs are well-established practices in several countries, including the United States, the United Kingdom, Canada, and Australia~\citep{clarke2009privacy, wright2012state}. They are also similar to Data Protection Impact Assessments (DPIAs) required in the European Union under Article 35 of the General Data Protection Regulation (GDPR) when processing activities are likely to pose a high risk. In this study, we refer to the DPIAs under the EU GDPR as part of the broader family of PIA approaches, despite some differences between them. Furthermore, the more recent EU AI Act mandates that deployers conduct DPIAs for certain high-risk AI systems under Article 26 and also prescribes Fundamental Rights Impact Assessments (FRIAs) under Article 27~\citep{AIACT2024}.

Over the years, significant research has been done on the development and implementation of PIA/DPIA methods~\citep{wairimu2024slr}. However, there are still gaps in understanding the state of the practice and the actual perceptions, experiences, and practices of privacy professionals (e.g., consultants, lawyers, officers, engineers) working with PIAs in organisations. In particular, existing research is scarce regarding the utilisation of PIAs/DPIAs after the enactment of the GDPR (``post-GDPR''), as mandatory legal requirements under Article 35 have propelled the field in both research and practice~\citep{iwaya2024privacy}. 

In this work, by PIA/DPIA ``in practice,'' we mean empirical experience, application, and practical use of PIAs in real-world settings. This also includes the interplay of emerging regulations, such as the EU AI Act, on PIA/DPIA practices across organisations. As such, this study aims to address this gap through a qualitative approach, drawing on 26 in-depth interviews with a diverse range of privacy professionals. For data analysis, the thematic analysis framework~\citep{braun2006using} was then employed to analyse the narratives and better understand how PIAs and DPIAs are conducted in practice, including whether the EU AI Act may influence their conduct.

As a result, this paper offers a multifaceted thematic framework of the state of practice in PIAs as its main contribution, comprising a rich and thick description of professionals' views and experiences that provides a comprehensive understanding of the topic. The descriptions capture participants' perspectives on PIA/DPIA-related activities, including their conduct and auditing, their views on the benefits and challenges, and the interplay between PIA/DPIA practice and the EU AI Act. Our contributions thus seek to help: (1) practitioners, by providing an empirical overview of how PIAs/DPIAs are implemented in real-world settings, including professionals' firsthand experiences and the operational challenges they face; and (2) researchers, by identifying critical gaps and suggesting avenues for further investigation on this topic.

With that in mind, the rest of the paper is organised as follows: Section \ref{related-work} reviews related work and identifies research gaps in the area. Section \ref{methods} highlights the study design, including data collection and analysis, as well as the ethical considerations. Section \ref{results} presents the results, including the thematic framework and key findings that compose its main themes and sub-themes. Section \ref{disc} presents the discussion, while Section \ref{threats-to-validity} explains the study's threats to validity and acknowledges limitations. Finally, Section \ref{conclusion} provides the study's main conclusions and outlook.


\section{Related Work}\label{related-work}
A recent systematic review has pointed to an evidence gap on the topic, as many studies on PIAs (and DPIAs) in practice yield contradictory results~\citep{iwaya2024privacy}.
Among the disputes, scholars have questioned the proven effectiveness of PIA methodologies~\citep{wadhwa2012privacy, friedewald2022dpias}, whether or not PIAs should be mandatory~\citep{wright2013introducing}, how to streamline PIA approaches~\citep{wright2013introducing} yet maintaining the assessment comprehensive and rigorous~\citep{clarke2016pia}, and the actual feasibility of conducting truly independent PIAs~\citep{edwards2012privacy, friedewald2022dpias}. 
Such issues underscore the need to balance the comprehensiveness and depth of assessments, adequately engage stakeholders (including data subjects), and tailor PIA approaches to each organisational context.

Significant research has also been done to identify existing PIA methodologies, compare them, and assess their maturity. Important contributions in the area are the theoretical studies that compared and evaluated the most common PIA methodologies, as in the works of \cite{vemou2018evaluation, vemou2019evaluating} and \cite{bisztray2019privacy}. Furthermore, a systematic review of the \emph{scientific} evidence on the evaluation and validation of PIA methodologies has been presented by \cite{wairimu2024slr}, revealing that such methods are seldom employed in practice, a finding corroborated by other studies ~\citep{iwaya2024privacy, hansen2025illuminating}.

In this study, we are particularly interested in better understanding the state of the practice through direct dialogue with privacy professionals working with PIAs and DPIAs in the real world.
For this reason, in Table~\ref{tab:related-work}, we summarise related work in the field that has addressed PIAs using qualitative methods, particularly interview-based studies.

\begin{table*}[ht]
\begin{center}
\footnotesize
\caption{\label{tab:related-work}Existing interview-based studies on the topic. Note: interview guides from other studies are found in the Appendices.}
\begin{tabular}{p{0.15\linewidth}p{0.1\linewidth}p{0.2\linewidth}p{0.075\linewidth}p{0.05\linewidth}p{0.275\linewidth}}
\hline
\textbf{Reference} & \textbf{Data Collection} & \textbf{Participants} & \textbf{Pop. Size} & \textbf{Interv. Guide} & \textbf{Other Comments} \\
\hline \hline
\cite{parks2011understanding} & Interviews, workshops and round tables & CIOs, CPOs, CMIOs, CEOs, HIPAA privacy compliance officers, and others & 17 / Unclear & Yes & Hospitals, healthcare associations, governments, EHR solutions, global research firms/consulting, United States. Grounded Theory. \\ \hline
\cite{brautigam2012pia} & Interviews & Privacy experts in the company & 15 & No & Finnish multinational (Nokia). \\ \hline
\cite{bamberger2012pia} & Interviews & 1 CPO, 3 former employees & 4 & No & Department of Homeland Security, United States. \\ \hline
\cite{wright2014integrating} & Survey, interviews & Unclear, survey respondents, senior managers & 148/12 & No & Government and private sector, United Kingdom. \\ \hline
\cite{pribadi2017regulatory} & Interviews & Project stakeholders & Unclear & Yes & IoT smart healthcare project, Indonesia. \\ \hline
\cite{van2017privacy} & Interviews & DPOs & 14 & Yes & Fourteen organizations across eight sectors, Netherlands. \\ \hline
\cite{ferra2020challenges} & Focus groups & 6 privacy consultants, 3 academics, 6 DPOs, 1 privacy activist & 16 & Yes & Europe. \\ \hline
\cite{vandercruysse2021dpia} & Interviews, survey & Data protection experts	& 16/18 & Yes & Stakeholders in smart cities, Belgium. \\ \hline
\cite{mckee2022pia} & Interviews, survey & Industry privacy professionals, members of standards initiatives, client representatives & Unclear & No & Medical device organisation, United States. \\ \hline
\end{tabular}
\end{center}
\end{table*}

As reported by \cite{iwaya2024privacy}, a significant turning point in PIA research was the entry into force of the EU GDPR in May 2018, which made DPIAs legally mandatory for high-risk systems.
However, as shown in Table~\ref{tab:related-work}, a significant portion of the related work predates the enactment of the GDPR.
Although significant insights were gathered into the privacy professionals' perspectives, a notable gap in evidence on post-GDPR practices remains.
Only three studies~\citep{ferra2020challenges, vandercruysse2021dpia, mckee2022pia} have begun to account for the actual effects of new GDPR requirements and privacy practices in organisations.

Still, although the work of \cite{ferra2020challenges} addresses DPIA practices and procedures, as well as issues and challenges, their workshop and focus groups (i.e., data collection) still took place four months before the GDPR came into effect.
And the work of \cite{vandercruysse2020typology} focuses on smart-city contexts, highlighting the clash of stakeholder interests rather than the concrete practices and experiences.
Finally, the work of \cite{mckee2022pia} is more limited, as it focuses on PIA practices within a single organisation.
Besides, it is also worth noting that we observed two types of methodological limitations across most related work, such as the absence of interview guides~\citep{brautigam2012pia, bamberger2012pia, wright2014integrating, mckee2022pia} and unclear population sizes~\citep{parks2011understanding, pribadi2017regulatory, mckee2022pia}.


In brief, related work in the area presents two main types of gaps: (a) evidence gaps in terms of persistent contradictory results coupled with a lack of post-GDPR observations; and (b) methodological gaps related to a narrow application-specific scope of study, or shortcomings of study design and methods used that can impact the validity and trustworthiness of their results. Notice that these gaps have already been reported in \cite{wairimu2024slr} and \cite{iwaya2024privacy} previously. Therefore, these research gaps motivated our work, which sought to address PIAs and DPIAs in practice more broadly in a post-GDPR context, employing rigorous qualitative methods.

\section{Methodology}\label{methods}

\subsection{Study Design}\label{design}
This study aimed to investigate professionals' perspectives on PIA/DPIA-related activities, including conducting, overseeing, auditing, and consulting. The professionals involved in the study included, among others, lawyers, privacy consultants and advisors, and Data Protection Officers (DPOs). To guide the study, the following research questions (RQs) were formulated:

\begin{itemize}
    \item \textbf{RQ1:} \textit{What are the professionals' perspectives of the PIAs in real-world practice?} \textbf{Objective:} To understand professionals' perspectives, experience, and understanding of PIAs/DPIAs, including the regulations and standards used in their work or organisation. 
    
    \item \textbf{RQ2:} \textit{How do professionals conduct and manage the various steps of the PIA process?} \textbf{Objective(s):} To understand how professionals organise and carry out PIAs/DPIAs in practice, including their integration into business processes; the allocation of responsibilities and stakeholder involvement; engagement with vendors or third parties; the methodologies and tools they use; risk assessment and selection of controls; conflict management; revision practices; perceived benefits and challenges; and the publication of PIA/DPIA results.

    \item \textbf{RQ3:} \textit{How do participants perceive the impact of new regulations on PIAs/DPIAs practices?} \textbf{Objective(s):} To understand professionals' perspectives on how new regulations, such as the EU AI Act, influence the conduct and management of PIAs/DPIAs, including whether and how Fundamental Rights Impact Assessments (FRIAs) are discussed or conducted within their organisations, and how professionals perceive Holistic Impact Assessments (HIAs).

\end{itemize}

To address the research questions, we opted for a series of semi-structured interviews as the data collection method, focusing on identifying organisations' core practices and procedures, as well as the main barriers and enablers to PIAs. Additionally, we sought to understand the professionals' perceptions of the utility and value of PIAs (i.e., positive and negative impressions). As such, we developed an interview guide (see Appendix \ref{interview-guide}) tailored to the specific group of professionals, enabling in-depth conversations on the topic. In addition to the interview guide, we also developed an ethical application package comprising the participant information sheet and consent form. 

We aimed to interview 18-30 people, using snowball sampling to reach theoretical saturation. The authors' professional networks were used to find and contact potential participants. The selection criteria were operationalised through screening questions included in the participant information sheet to verify (i) participants' organisational roles, and (ii) their experience in conducting, overseeing, auditing, or consulting on PIAs. These questions ensured that participants possessed relevant professional expertise and direct involvement with PIA processes.


\subsection{Data Collection}
The data collection phase started in February 2025 and ended in April 2025. During this period, we contacted individuals in our professional networks (e.g., lawyers, privacy engineers, DPOs, and consultants) via email invitations to participate in the study. Additionally, we employed snowball sampling~\citep{parker2019snowball} by asking initial contacts to refer us to additional participants who met the study criteria. Upon agreeing to participate, participants were sent the participant information sheet and the consent form. Furthermore, we requested that they submit their consent forms before scheduling the interview. By the end of the study period, we had sampled and interviewed 26 participants with experience in PIA-related activities.

Interviews were scheduled and conducted via Zoom, each lasting between 60 and 90 minutes. The interviews were conducted by 2-3 interviewers, with one leading the interview and the others taking notes and asking additional questions when needed. The interview followed a semi-structured format based on a predefined interview guide, while allowing flexibility to explore emerging topics and ask follow-up questions. In total, the 26 interviews yielded approximately 27 hours of audio recordings, with interviews averaging slightly over 1 hour each.

\subsection{Data Analysis} 
After the study, the interviews were transcribed by three researchers, with all identifying information removed, including participant and organisation names. During the transcribing process, the researchers familiarised themselves with the data by rewatching interviews and reading the transcripts. The interviews were transcribed verbatim to preserve accuracy and retain the original meaning. However, filler words (e.g., ``um'', ``uh'', etc.) were removed to enhance readability. Furthermore, the removal of filler words was intended to enhance readability as we adopted Interviewee Transcript Reviews (ITRs), which enabled participants to review their interview transcripts for accuracy, correct errors, and provide clarifications~\citep{hagens2009interviewee}. All participants were asked to validate that the transcripts accurately reflected the discussions and to identify any key points that may have been omitted or misrepresented.

Following transcript preparation and validation, the interview data were analysed using thematic analysis, as described by~\citep{braun2006using, braun2019reflecting}. In particular, the analysis primarily followed an inductive approach (i.e., data-driven), as themes were developed from participants' accounts and grounded in the data. However, the analysis was also informed by the study's research questions, introducing a deductive element to the analytic process. To initiate the coding process, one transcript (i.e., P4) was randomly selected for preliminary analysis. All three authors independently imported the transcript into their NVivo 14 software\footnote{NVivo 14 Qualitative Data Analysis Software (\url{https://lumivero.com/products/nvivo/})} and coded the transcript using a qualitative coding approach. In addition, the authors independently generated preliminary codes from the transcript and subsequently met to discuss their interpretations. These discussions informed the development of an initial set of codes that guided the subsequent stages of analysis. 

Thereafter, the first author coded the remaining 25 transcripts. During this process, the development of new codes and sub-codes was continuously documented and discussed among the authors. Theoretical saturation was observed at P19, as few or no codes emerged, with only minor refinements through additional sub-codes. After all transcripts had been coded, the first author defined the initial themes and sub-themes, which were subsequently reviewed and refined collaboratively during project meetings. The final thematic framework was iteratively refined through collaborative discussions among the authors, enabling reflection on different interpretations and ensuring that the themes meaningfully captured patterns across the dataset. A rich, multifaceted, and thick description of the resulting thematic framework is thus produced, covering not only the most predominant perspectives but also nuanced observations from participants.

\subsection{Ethical Considerations}\label{ethical}  
Since our study involved human subjects, we ensured compliance with research ethics at the [\textit{University Name Anonymised}]. We submitted the ethical application, including the participant information sheet, consent form, project plan, and interview guide, to the local ethics committee for evaluation. We received approval to conduct the study from the university's ethics advisor (registration number [\textit{XX 2024-XXX}]). 
Participation in the study was entirely voluntary, and informed consent was obtained before the interviews, including video and audio recordings. Participants were assured that transcripts, video, and audio would be processed in a manner to prevent unauthorised access. We restated the study's objective to the participants during the interview session and invited any additional questions before commencing with the study.

\section{Results}\label{results}
This section presents the results of the study based on professionals' perspectives on PIA-related activities, such as conducting, auditing, and consulting. 
First, we provide an overview of participant demographics, the number of PIAs/DPIAs they conducted, and the main privacy laws, regulations, guidelines, and standards that inform their PIA/DPIA practices, which answers RQ1. We then present the key findings of the study, organised into a thematic framework.

\subsection{Participants Characteristics}
\begin{figure}
  \centering
    \includegraphics[width=1.0\columnwidth]{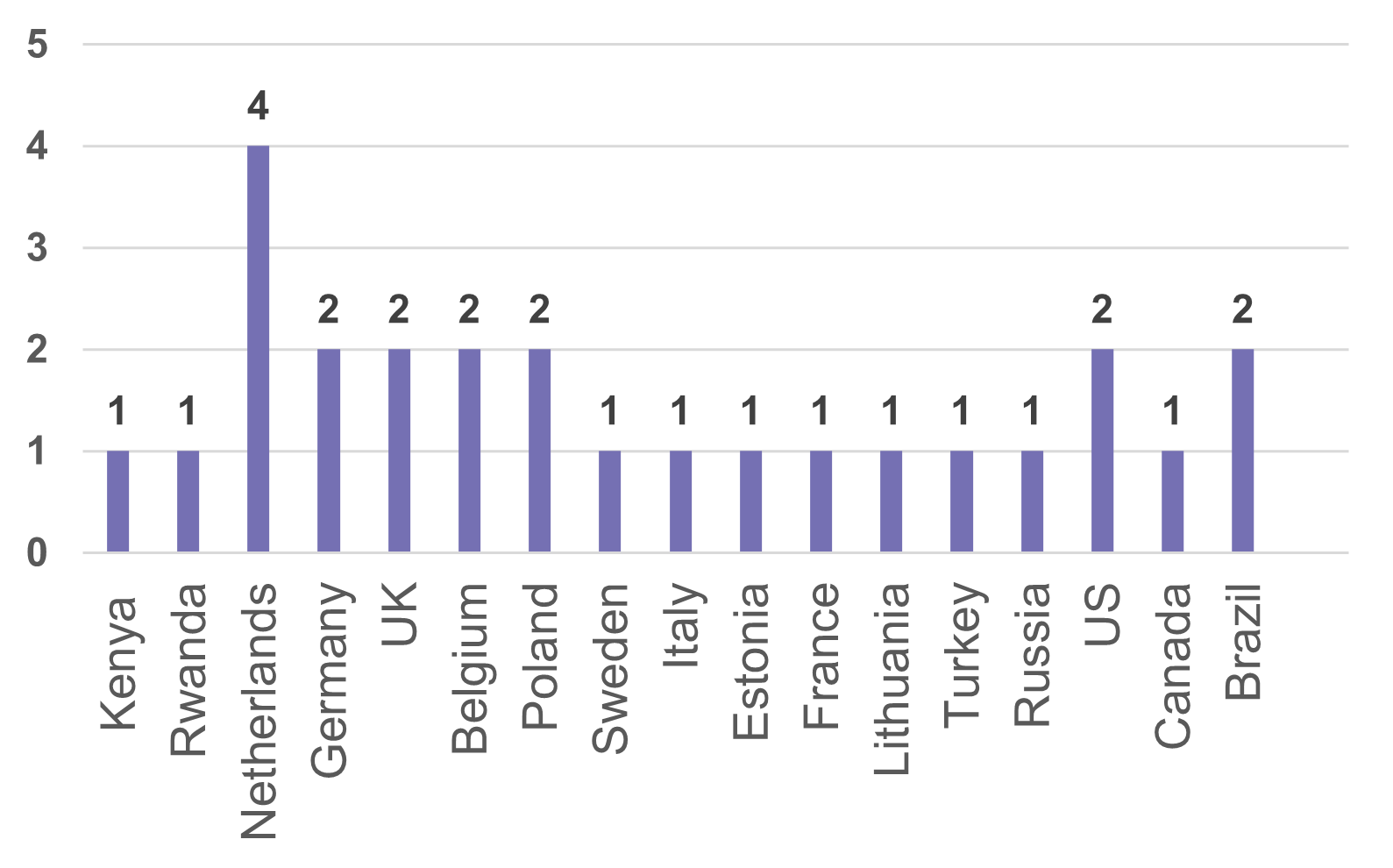}
    \caption{Geographical distribution of interview participants.}\label{fig:countries}
\end{figure}

This study involved 26 participants from 17 countries across Europe, North America, South America, Africa, and Asia, as illustrated in Figure \ref{fig:countries}. The study had the highest number of participants from the Netherlands, followed by Germany, the UK, Belgium, Poland, the US, and Brazil. Other countries followed with one participant each. Participants from European countries ($n=19$, 73\%) (i.e., counting Russia and Turkey) formed the most representative group.

Table \ref{tab:participants} outlines the role and years of experience of the participants, including the number of PIAs/DPIAs they conduct or have conducted in practice. The participants represented diverse and multidisciplinary professional backgrounds spanning privacy, data protection, consulting, law, governance, and academia. Most participants worked primarily in privacy-related roles, including DPOs, privacy advisers, GDPR specialists, privacy consultants, and privacy engineers, indicating strong expertise in data protection and regulatory compliance. Their experiences came from industry and research institutions, including the financial sector, the data protection regulator, consultancy firms, transport, the automotive industry, gaming \& e-sports, healthcare, the software industry, universities, and consumer advocacy and standardisation. 

Participants ranged in experience from 4 to 29 years. Most participants ($n=9$, 34.6\%) had between 6 and 10 years of experience, while the fewest participants ($n=4$, 15.4\%) had 5 or fewer years of experience. A substantial set of participants ($n=7$, 26.9\%) had between 11 and 20 years of experience, followed by 23.1\% participants ($n=6$) with 21-30 years of experience. Over the years, participants have contributed to the field of data protection in several ways. For example, P04 contributed to the standardisation initiatives (e.g., NIST, ISO) in the beginning and developed a method for AI assessment; P21 \emph{``participated in many of the Working Party 29 opinions, including the opinion on DPIA''}, and P15 \emph{``served as an expert at the Data Protection Unit of the Council of Europe.''}

Furthermore, Table \ref{tab:participants} outlines the number of PIAs/DPIAs conducted by participants both within and outside their organisations. It is important to note that the numbers are approximate. The participants reported conducting multiple PIAs/DPIAs as well as ROPAs (Register of Processing Activities). Taking some examples into consideration, P10 stated that they review EU DPIAs approximately once a month and have conducted one to two Canadian PIAs in other years. However, at the time of the interview, (P10) estimated that approximately ten PIAs would be conducted that year. (P21) reported conducting approximately four \emph{``serious''} DPIAs that they regarded as particularly noteworthy, as well as $\approx 20$ to $30$ assessments of a less substantial nature. They noted that the complexity of the assessments depended on the specific processing activities evaluated. 

\begin{table*}
  \centering
  \small
  \caption{Participants' demographic details and the number of PIAs/DPIAs conducted in practice. Note: Participant ID (P. ID) and Experience (Exp.).}
    \begin{tabular}{lp{0.4\textwidth}llp{0.275\textwidth}}
    \hline
    \textbf{P. ID}&\textbf{Professional Title} & \textbf{Country} & \textbf{Exp.} & \textbf{\#PIAs/DPIAs}  \\
    \hline\hline
    P01 & Consultant, Standardisation Committee Member & Lithuania & $\approx 7$ & 2--3/month
    \\  \midrule
    P02 & Project Manager, Consultant & Belgium & $\approx 20$ &  >100
    \\ \midrule
    P03 & Information Security Risk Management, Data Protection Expert & Germany & $\approx 20$ & -
    \\ \midrule
    P04 & Consultant Advisor & Netherlands & >20 & >100
    \\ \midrule
    P05 & Senior Researcher, Group Leader & Germany &  >25 & DPIA testing with 12--15 companies
    \\ \midrule
    P06 & DP, GDPR specialist & Sweden & $\approx 29$ & 3
    \\ \midrule
    P07 & Privacy Engineer & US & $\approx 25$ & >300 
    \\ \midrule
    P08 & Privacy Advisor & UK & $\approx 7$ & >50
    \\ \midrule
    P09 & Consultant Advisor & UK & $\approx 20$ & $\approx 30$--$40$
    \\ \midrule
    P10 & Lawyer, Consultant & Canada & $\approx 27$ & EU DPIA: $\approx 1$/month; Canada PIA: 1--2/year (2025: $\approx 10$)
    \\ \midrule
    P11 & Privacy Lawyer & Italy & $\approx 10$ & $\approx 30$
    \\ \midrule
    P12 & Head of Privacy \& AI Governance, Private Consultant, Academic & Poland & >10 & 100--150 
    \\ \midrule
    P13 & Risk and Compliance Expert & Estonia & $\approx 5$ & 2
    \\ \midrule
    P14 & Privacy Consultant, Adjunct Professor & US &  >20 & <50
    \\ \midrule
    P15 & Lead Technology Policy \& AI Governance, Advisor & Rwanda & $\approx 8$ & >40 (across multiple sectors)
    \\ \midrule
    P16 & DPO & Russia & $\approx 6$ & $\approx 1000$
    \\ \midrule
    P17 & External DPO, Consultant & France & $\approx 10$ & 10--15+
    \\ \midrule
    P18 & Group Data Protection Specialist & Turkey & $\approx 16$ & >10
    \\ \midrule
    P19 & Lecturer, Consultant & Netherlands & $\approx 6 - 7$ & $\approx 40$--$50$
    \\ \midrule
    P20 & Senior Privacy Adviser, External DPO & Netherlands & $\approx 19$ & $\approx 50$
    \\ \midrule
    P21 & Researcher & Belgium & $\approx 4$ & DPIAs: $\approx 4$ major; $\approx 20$--$30$ minor
    \\ \midrule
    P22 & Lawyer & Kenya & 5 & $ \geq 12$
    \\ \midrule
    P23 & Auditor & Poland & 5 & >12 
    \\ \midrule
    P24 & DPO & Netherlands & >7 & $ \approx 60$
    \\ \midrule
    P25 & Global Privacy Program Manager & Brazil & 7 & >500
    \\ \midrule
    P26 & Data Governance Manager & Brazil & 7 & $\approx 5000$--$6000$ ROPAs
    \\ \hline
    \end{tabular}%
  \label{tab:participants}%
\end{table*}

\subsection{Privacy Laws, Guidelines, and Standards}
\begin{figure}
  \centering
    \includegraphics[width=1.0\columnwidth]{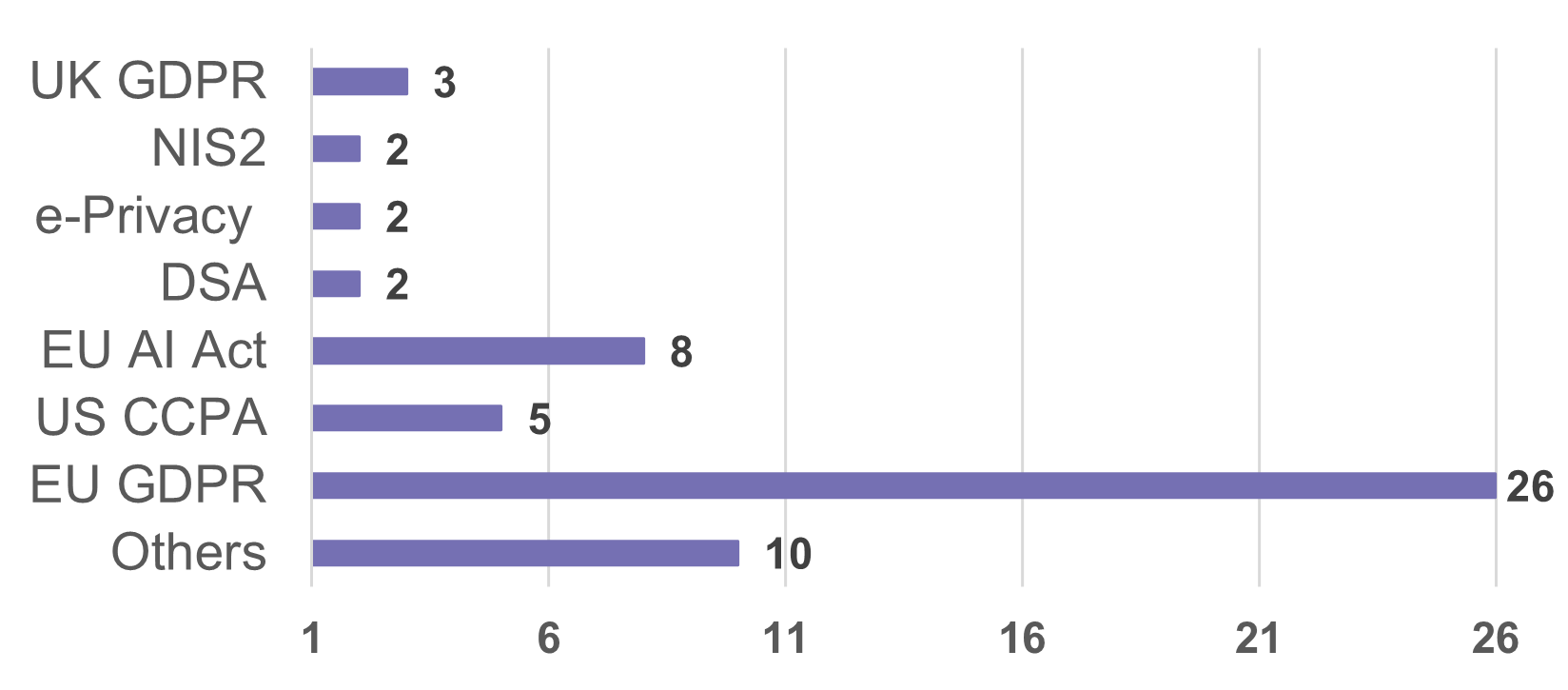}
    \caption{Reported privacy laws and regulations.}\label{fig:regulations}
\end{figure}

Figure \ref{fig:regulations} depicts the privacy laws and regulations that participants mentioned as the most important in their organisations, or with which they had worked to inform their PIA practices. While several regulations were mentioned, we focused on privacy laws and regulations that were mentioned at least twice. Other laws and regulations mentioned only once were: Digital Operational Resilience Act (DORA), UN Regulation No. 155 on cybersecurity management systems, Digital Markets Act (DMA), electronic IDentification, Authentication and trust Services (eIDAS), Cyber Resilience Act, EU Digital Strategy, EU Data Act, Malabo Convention, European Health Data Space (EHDS), and the Charter of Fundamental Rights of the European Union. 

As expected, most participants reported using or relying on the EU GDPR. While established in the EU, its reference particularly outside the EU suggests it's a common reference framework across participants' professional environments. Specifically, participants reported that EU GDPR is \emph{``kind of a gold standard''}(P14) ,\emph{``globally accepted''} (P08), and recognized by \emph{``many DPOs worldwide''} (P16). Other regulations that followed include the EU AI Act, US CCPA, DSA, e-Privacy, and NIS2. In addition to working with the EU GDPR, participants also mentioned working with or referring to national laws and guidelines. Some of the national laws mentioned include PIPEDA, LGDP, UK DPA, Lithuania Data Protection Law, Kenyan DPA, Dutch telecommunications law (implementing the ePrivacy Directive), as well as sector-specific laws like HIPAA. Guidelines reported also varied across participants, with the most common being those from the European Data Protection Board (EDPB), the French National Commission on Informatics and Liberty (CNIL), the Spanish Data Protection Agency (AEPD), and the UK Information Commissioner's Office (ICO).

\begin{figure}
  \centering
    \includegraphics[width=1.0\columnwidth]{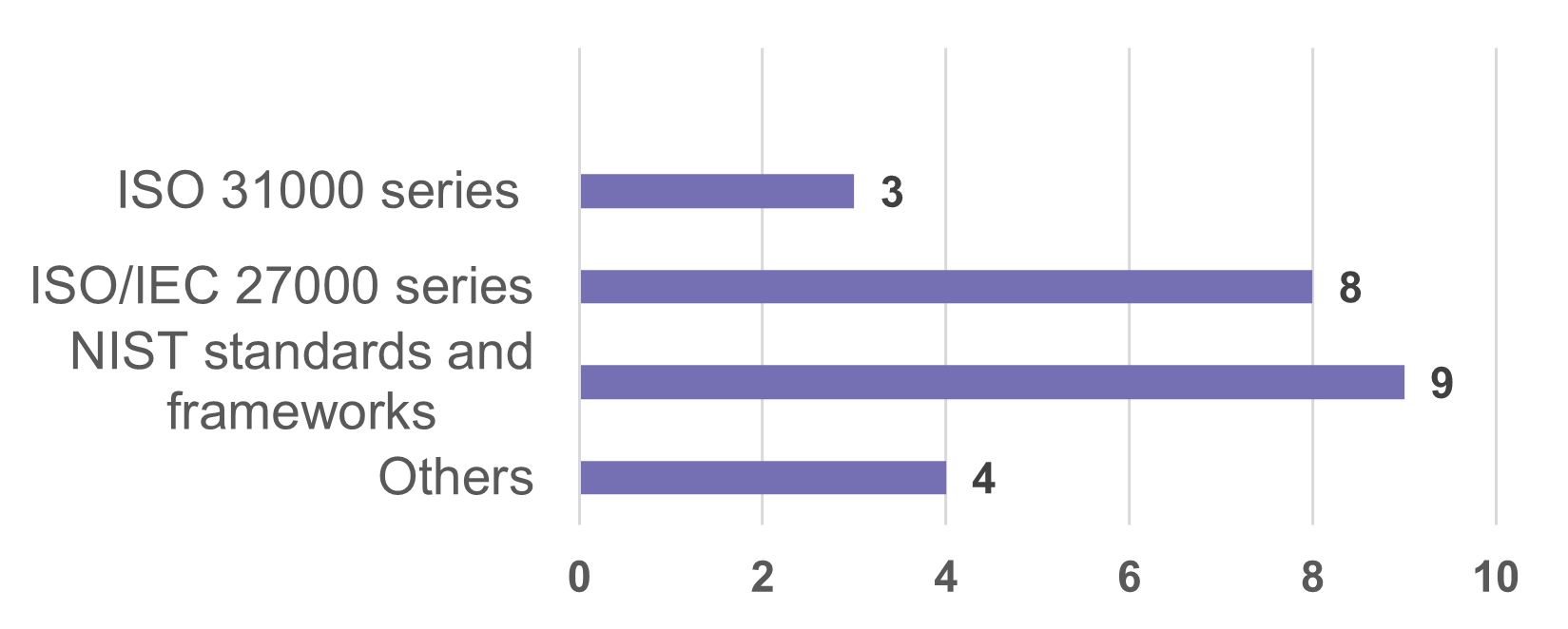}
    \caption{Reported privacy standards.}\label{fig:standards}
\end{figure}

In addition to the privacy laws and regulations, participants were asked whether they used privacy standards to inform their PIA practices. Figure~\ref{fig:standards} illustrates the privacy standards reported by participants. As observed, nine participants referred to National Institute of Standards and Technology (NIST) standards and frameworks. While some participants specified particular publications (such as the NIST Cybersecurity Framework, NIST Privacy Framework, and NIST Risk Management), others referred to NIST more generally without identifying a specific standard. Similarly, this was observed with the International Organization for Standardization and the International Electrotechnical Commission (ISO/IEC) 27000 series. Several participants specified particular ISO standards (such as ISO 27001, ISO 27002, ISO 27003, ISO 27004, ISO 27005, ISO 27701, and ISO 29134), and a few referred to ISO standards more generally. This, however, was different from the ISO 31000 series, in which specific standards were reported. These are ISO 31000 and ISO 31010. Other standards that were mentioned only once were: the Payment Card Industry Data Security Standard (PCI DSS), the General Health Information Systems Security Policy (PGSSI-S), the System and Organization Controls (SOC 1 \& 2), and the Singapore framework (in the context of AI governance).

\subsection{Key Findings}\label{findings}
\begin{figure*}
  \centering
    \includegraphics[width=1\textwidth]{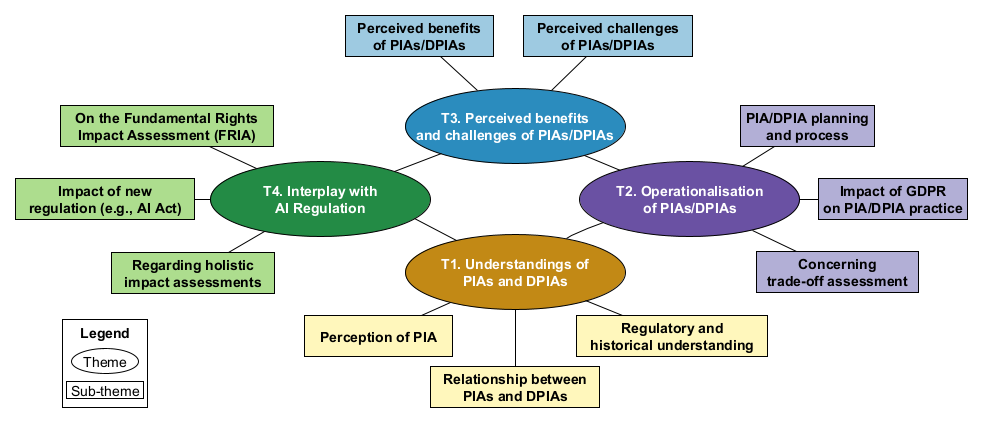}
    \caption{Overview of the study's thematic framework.}\label{fig:thematic_map}
\end{figure*}

In the following section, we present key findings, which we organise as a narrative to address the research questions in subsection~\ref{design} based on the thematic framework illustrated in Figure~\ref{fig:thematic_map}. The thematic framework was primarily developed using an inductive (i.e., data-driven) approach; however, we also drew on a deductive approach to ensure that the analysis remained aligned with the research questions. It is important to note that this overview of the thematic framework in Figure~\ref{fig:thematic_map} is simplified to reduce complexity and present the data in a structured manner, while providing an overview of the themes and sub-themes that aim to tell the narrative of professionals' perspectives and PIAs' practices in organisations.

A brief description of each theme is provided below:
\begin{itemize}
  \item[T1.] \emph{Understandings of PIAs and DPIAs:} This theme captures how participants understand PIAs and DPIAs, including whether they perceive differences between the two exercises. 
  \item[T2.] \emph{Operationalisation of PIAs/DPIAs:} This theme describes how the participants plan and conduct PIAs and DPIAs. In addition, it captures the impact of the EU GDPR on DPIA practice and covers how they assess trade-offs. 
 \item[T3.] \emph{Perceived benefits and challenges of PIAs/DPIAs:} This theme explores participants' perceptions of the benefits and challenges associated with conducting PIAs and DPIAs. It encompasses both direct responses to explicit questions about benefits and challenges and related perspectives that emerged organically during discussions of other topics.
 \item[T4.] \emph{Interplay with AI regulation:} This theme encompasses discussions on the potential impact of emerging regulations, such as the EU AI Act, on future PIA/DPIA practices. It included participants' familiarity and experiences with Fundamental Rights Impact Assessments (FRIAs), as well as their views on combining multiple assessments into a unified, holistic impact assessment (HIA).
\end{itemize}

In the following subsections, we present and discuss the themes and sub-themes identified through the thematic analysis, supported by illustrative quotes from the participants. 

\subsubsection{T1. Understandings of PIAs and DPIAs}
\begin{figure}
  \centering
    \includegraphics[width=1.0\columnwidth]{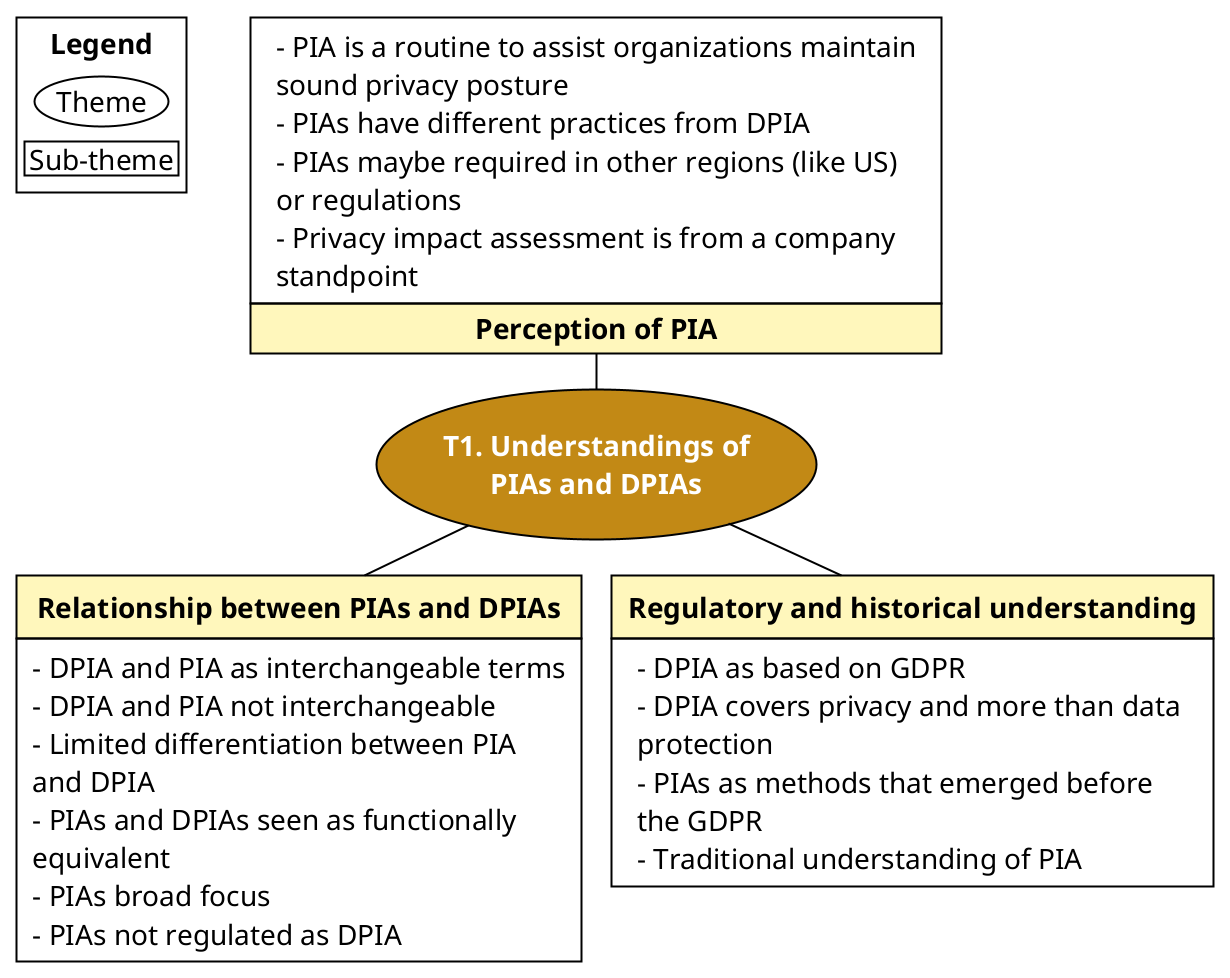}
    \caption{Overview of theme 1 and its sub-themes.}\label{fig:theme_1}
\end{figure}

Figure~\ref{fig:theme_1} illustrates the overarching theme, its sub-themes, and the summarised codes identified for each sub-theme. To provide a deeper understanding of participants' conceptualisations and answer RQ1, each sub-theme is discussed in turn below.

\subsubsection*{Sub-theme T1.1: Perceptions of PIAs}
In this sub-theme, professionals' perceptions of PIAs are presented, including perceived usefulness, as well as their conceptualisations of the term PIA. 
Regarding the usefulness of PIAs, participants described them as organisational risk-management tools (P02, P16) rather than as mechanisms for protecting data subjects' interests. Several participants emphasised that a PIA is \emph{``focused on the impact on the organisation''} (P24), and is \emph{``generally meant to just maintain a proper and sound privacy posture''} (P22). These suggest that the participants view a PIA as conducted from a company's standpoint rather than a data subject's. 

More generally, P22 and P16 articulated their conceptualisation and understanding of a PIA. Their accounts suggest that they understood a PIA in two general ways. First, as a routine performed on a rolling or continuous basis whenever an organisation changes a process or tweaks a service offering, it is carried out regardless of whether the processing activity presents a high risk, with the primary purpose of helping the organisation maintain a proper and sound privacy posture (P22). Second, as a generally voluntary control that a company, controller, or DPO may conduct. Its goal is to identify the risks the company must address and support decisions on how those risks should be managed, including whether to accept or transfer them (P16).

Furthermore, it was observed that PIAs \emph{``have different structures''} (P21) compared to the requirements set out in Article 35(7) of the GDPR~\citep{GDPR2016}. This suggests that, while DPIAs may employ different methodologies, they are guided by a common set of legal requirements, whereas PIAs may vary substantially in their structures and organisational practices.

Participants also associated the term PIAs with jurisdictions outside the European Union (EU). Given that PIAs have their origins in Anglophone countries, particularly Australia, Canada, New Zealand, and the US~\citep{clarke2009privacy, clarke2011evaluation}, participants perceived PIA as a term primarily used \emph{``outside the European Union''} (P19), particularly in the US (P17, P18).


\subsubsection*{Sub-theme T1.2: Relationship Between PIAs and DPIAs}
Participants also discussed their understanding of PIAs, particularly their relationship to DPIAs, including their functionality, terminology, and the differences between them. Regarding terminology, participants expressed mixed views on whether the terms should be used interchangeably. On the one hand, it was noted that PIA and DPIA are sometimes \emph{``used interchangeably''} (P15). On the other hand, several participants argued that PIA and DPIA \emph{``should not be synonyms''} (P19, P24), explaining that \emph{``privacy is different from data protection''} (P19).

Given this difference, where privacy was viewed differently from data protection, and where PIA and DPIA were expressed as dissimilar, further views were expressed. In particular, discussions about whether PIAs assess data protection, encompass data protection, or are broader than DPIAs were noted. For instance, P13 explained that \emph{``in the context of DPIA and PIA, PIA (...) encompasses [the] data protection part''}. Similarly, P24 stated that \emph{``a privacy impact assessment is broader because privacy is not the same thing as data protection''}. These two responses suggest that, indeed, PIAs are more expansive and they tend to cover not only data protection, but privacy in general. In other words, PIAs look \emph{``at privacy in the broader sense, unlike a DPIA that has specific triggers''} (P15). These triggers are defined in the GDPR and apply when data processing is likely to pose high risks or when other criteria are met~\cite{wp2017guidelines}.

Furthermore, other participants described the differences between the two exercises. P25 argued that they preferred the \emph{``term privacy impact assessment because it allows (...) to check the data protection requirements, but also how it's going to impact [data subjects] in their private sphere''}. This distinction reflects the view that privacy encompasses broader concepts, such as autonomy and other dimensions of privacy that relate to data subjects~\citep{Finn2013}, while also accounting for data protection requirements. In contrast, two participants argued that DPIAs are similarly broad in scope. P13 stated that privacy can also be assessed through a DPIA. At the same time, P20 explained that a DPIA is concerned not only with protecting personal data but also with safeguarding broader fundamental rights and freedoms that may be affected by data processing.

Participants also identified other differences between the two assessments. Unlike a DPIA, which is specifically regulated under the GDPR, a PIA was described as an unregulated assessment. As P23 explained, a \emph{``PIA can be conducted all over the world based on international standards''} (P23) and as such, not  \emph{``required by law''} (P21). Similarly, P16 argued that PIAs are not regulated by legislation \emph{``in most countries and most jurisdictions''}, unlike DPIAs, which are subject to specific legal requirements. In addition, P19 noted that, unlike DPIAs, PIAs are not part of the GDPR.

However, despite this differentiation, several participants described PIAs and DPIAs as functionally equivalent. For instance, P21 stated that PIAs and DPIAs \emph{``have the same logic''} even though the terminology differs. Similarly, P10 argued that the purpose of a DPIA and a PIA is essentially the same, as both involve analysing data processing activities to ensure that personal data is collected for legitimate purposes, shared and disclosed appropriately, supported by legal authority, and adequately safeguarded. Likewise, P12 argued that PIAs and DPIAs differ primarily in their terminology and jurisdiction, as discussed earlier. Nevertheless, the participant emphasised that both assessments fundamentally involve evaluating how a processing activity affects privacy and identifying measures to mitigate privacy risks. Consistent with this view, P08 suggested that the concept of \emph{`` privacy by design and default, which will come under a privacy impact assessment''} also applies under a DPIA, although under the GDPR term data protection by design and by default (Art. 25), reflecting the same underlying principle.

Following P08's perspective, P05 described the assessment as more than a mechanism for GDPR compliance, arguing that a PIA or DPIA should serve as a practical tool that supports privacy by design. The participant suggested that embedding privacy considerations throughout the design process could improve the quality of products and services while increasing stakeholder confidence that privacy had been adequately addressed.


\subsubsection*{Sub-theme T1.3: Regulatory and Historical Understanding}
Participants further described their understanding of PIAs and DPIAs from both a regulatory and historical perspective. They generally viewed PIA as \emph{``an old-fashioned word''}(P20) or a term that \emph{``existed before''} (P15) the GDPR, describing it as an assessment that was considered \emph{``good practice that was developed since the 80s''} (P21). As discussed earlier, participants associated PIAs with jurisdictions outside the EU, specifically the US. In a later discussion, one participant explained that, traditionally, PIAs were required for US government agencies (P14). By contrast, private companies were not generally required to conduct PIAs and often did so voluntarily. As a result, the participant described company PIAs as \emph{``much looser''} than those used in government. Nevertheless, this account suggests that PIAs were indeed regarded as good practice, as companies adopted them despite the absence of a legal requirement.

Given that PIAs were classified through a traditional lens, participants also provided a distinction between PIAs and DPIAs from a regulatory perspective. Generally, participants described a DPIA as a term associated with the EU GDPR (P01, P11, P12, P13, P14, P15, P16, P17, P19, P23) and hence, having a legal aspect. 
Participants distinguished between PIAs and DPIAs based on their perceived legal status. While some participants noted that PIAs are not necessarily a legal requirement in most countries, several described DPIAs as mandatory assessments under the EU GDPR (P01, P04, P16, P18, P19, P21, P22, P23). However, participants clarified that a DPIA is only \emph{``mandatory in some cases''} (P18), based on specific triggers \emph{``defined by the GDPR or other laws or even like the ones the data protection authorities or the EDPB have kind of curated''} (P15). The other laws in this case could be national, as \emph{``each member state could add additional cases or situations when the DPIA is mandatory''} (P01).

Following this, participants described the purpose of a DPIA and distinguished it from a PIA. While some participants emphasised that PIAs are conducted from an organisation's perspective, others argued that DPIAs are conducted from the perspective of the data subjects. Specifically, participants explained that a DPIA focuses on \emph{``trying to find out what the risks are to the fundamental rights and freedoms of the people whose data you are going to be processing''} (P24), rather than on risks to organisations. However, it was also noted that a DPIA is only required when there is a sufficient level of risk, with one participant describing this threshold as \emph{``when there is suspicion that there is a high risk (...) to processing activities that are involved''} (P22).  


\hfill \break 
\noindent\fbox{
    \begin{minipage}{.95\linewidth}
        \setlength\parskip{1em}
        \textit{\textbf{Key Takeaways:}\\
        $\rightarrow$ PIAs were viewed as a pre-GDPR practice commonly associated with jurisdictions such as the US and the UK, whereas DPIAs were associated with the GDPR, characterised by specific legal requirements.  \\
        $\rightarrow$ PIAs were viewed as assessments centred on organisational risks, while DPIAs were understood to focus on risks to data subjects. Some participants, however, argued that PIAs also address broader privacy concerns and impacts on data subjects. \\
        $\rightarrow$ Despite differing views on their terminology and scope, many participants regarded PIAs and DPIAs as functionally equivalent assessments that share the same underlying purpose.
        } 
    \end{minipage}
}

\subsubsection{T2. Operationalisation of PIAs/DPIAs}
\begin{figure}
  \centering
    \includegraphics[width=1.0\columnwidth]{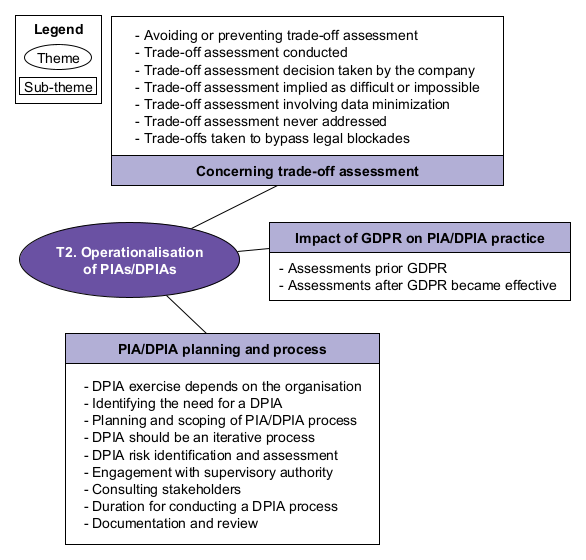}
    \caption{Overview of theme 2 and its sub-themes.}\label{fig:theme_2}
\end{figure}

Figure \ref{fig:theme_2} illustrates the theme and sub-themes, including summarised codes for each sub-theme, which partly address RQ1 and fully address RQ2.

\subsubsection*{Sub-theme T2.1: Impact of GDPR on PIA/DPIA practice}
Before GDPR introduced DPIAs, PIAs were already performed by some participants (P04, P07, P11), as they had \emph{``been in existence for quite some time''} (P03). Still, they lacked standardisation; that is, there was \emph{``no standard way''} of conducting PIAs (P07). Following GDPR, participants reported experiencing a transitional period characterised by an evolution to DPIA (P04), limited guidance (e.g., lack of guidance and documentation; P05, P08, P11, P14, P21), scarce practical experience (P05), and a lack of methodologies (P01). Considering the latter, the participant reported relying on emerging methodologies, such as the CNIL methodology. At the same time, others developed their own DPIA methodology that was \emph{``influenced by the French Data Protection Agency methodology''} (P21), i.e., CNIL. Given the early stage of DPIAs, P08 expressed that \emph{``everybody was just learning from one another''} and, after a short while, they started taking courses and earning certifications to update their knowledge of what one was supposed to be doing. As such, it can be argued that, while the GDPR affected impact assessment practice, participants (as well as organisations) had to develop much of the practical implementation themselves. This was still reflected in the DPIA planning and process, as we shall discuss in the following theme.

\subsubsection*{Sub-theme T2.2: PIA/DPIA Planning and Process} 
\begin{figure}
  \centering
    \includegraphics[width=1.0\columnwidth]{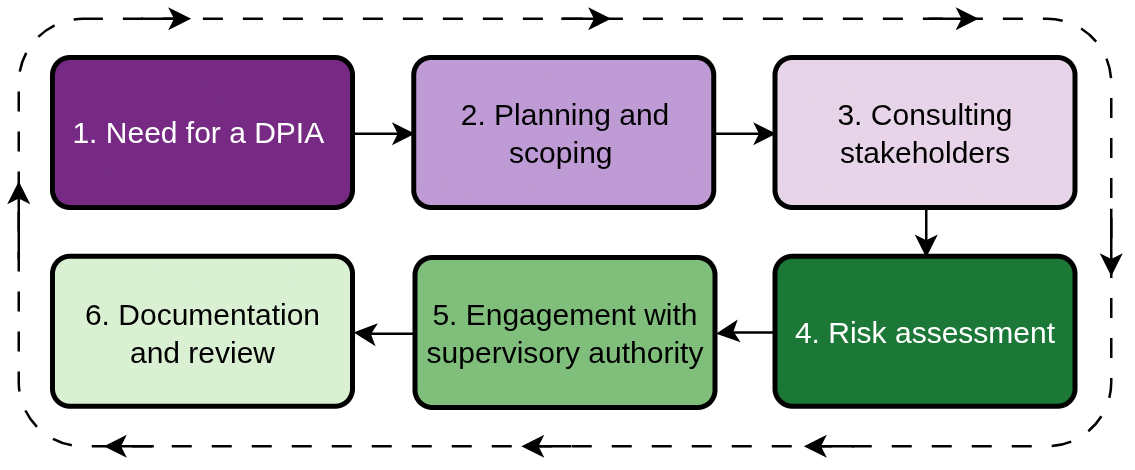}
    \caption{Generalised activities depicting how participants plan and conduct PIAs/DPIAs. The boundary indicates an iterative process.}\label{fig:process}
\end{figure}

Having developed an understanding of the impact of GDPR, we sought to answer RQ2 by exploring how DPIAs are conducted in practice. Participants' accounts revealed seven activities associated with their planning and conducting of a DPIA. These activities are illustrated in Figure~\ref{fig:process}, which provides a generalised representation of a DPIA process. Importantly, not all participants reported the same activities or described them in the same order. At the same time, some participants expressed that certain activities were left for the organisation to complete, highlighting variations in how DPIAs are operationalised in practice. Indeed, several participants described that a DPIA exercise depends on an organisation. For example, P14 explained that a DPIA \emph{``varies widely by organisation''}, while P09 described that the exercise can be approached in various ways.
In the following sections, we analyse the activities in the order shown in Figure \ref{fig:process}.

\paragraph{\textbf{Need for a DPIA:}} Participants explained that, before conducting a DPIA, it was necessary to determine whether one was required. As such, 13 (50\%) participants reported performing a threshold assessment, while 9 (34.6\%) assessed triggers. This initial assessment allowed them to determine whether the proposed processing activity presented a sufficiently high risk to the data subject's rights and freedoms to warrant a DPIA. For example, P01 and P26 described that they prepared a ROPA (Records of Processing Activities) to ascertain whether a DPIA was necessary. As such, where one or more triggers were identified, the processing activity was subject to a ``full-blown'' DPIA.

However, regarding threshold assessments, two participants expressed some criticism. P07 noted that if an assessor does not understand the process well enough, they could fill out a threshold assessment incorrectly. As such, the assessment would indicate that there are few or no privacy risks, rendering a detailed DPIA unnecessary. Similarly, P14 cautioned against relying on broad binary answers in threshold assessment. For example, asking whether the project uses personal data may enable respondents to answer no, and thereby avoid a full DPIA, even when the project involves information that constitutes personal data. Instead, the participant emphasised that more specific questions regarding data categories should be asked, such as biometric data and names. Such questions can make it more difficult to overlook relevant personal data, thus determining whether a full-fledged DPIA is necessary. 

\paragraph{\textbf{Planning and scoping:}} 
This activity encompassed several sub-activities. Participants discussed their understanding of the scope of the DPIA process, responsibilities throughout the DPIA process, the integration of DPIAs into broader processes, the description of processing activities, DPIA design approaches and methodologies, and the timing of DPIAs within a project. Regarding their understanding of the scope of the DPIA process, five (19.2\%) participants discussed how and why they needed to understand it. P10, for example, explained that they meet with clients \emph{``to get a sense or a flavour of what they are analysing, what they are implementing, [and] whether it [is] a new technology or a new program''}. P24 further described what they \emph{``would do is indeed organise a kickoff session to set the exact scope''}. 

Following the understanding of the scope, another sub-activity involved describing the processing activities. Thirteen (50\%) participants discussed how they establish which personal data is involved, what processing is being assessed, who is affected, and how to identify or get such information. P11, P13, P15, and P16 explained that they would gather such data from relevant people through means such as interviews or questionnaires, depending \emph{``on the level of complexity (...) of the processing activity''} (P11). In addition, P02 and P18 explained that they use data flow diagrams to depict how the data moves through the organisation. Specifically, P02 stated that the initial thing they do \emph{``is to design the data flow in the organisation and to show all these points where the data is going''}, while P18 also expressed that they \emph{``want to show how the data will flow, how it starts, and how it ends''}.

Participants further discussed, as part of their planning, the methodologies they use to conduct the assessment. Discussions of these methodologies revolved around the development and use of bespoke methodologies, the use of DPIA methodologies inspired or created by several supervisory authorities, the use of guidance from supervisory authorities, and the need for a harmonised DPIA. Generally, these discussions painted a picture of a lack of standardised DPIA methodology, templates, and guidance, as expressed by P10, P11, P19, P04, P02, and P01. For example, 14 (54.8\%) participants explained that they developed and use bespoke methodologies for internal use within an organisation or when consulting for other companies, while three participants (P02, P04, and P19) noted that the methodology they used to conduct a DPIA depended on the specific organisation. Hence, two participants (P12, P17) called for standardisation. For example, P17 called for \emph{``a single methodology across the EU''}, including a single list of processing activities requiring a DPIA, while P12 emphasised the need for a unified and clearly explained template. Nevertheless, not all participants conduct DPIAs; one participant stated that they prefer to conduct privacy threat modelling because PIAs are \emph{``just not substantial enough''} because \emph{``they don't necessarily get to the real problem''} (P07). Also, 24 (92.3\%) participants described whether the DPIAs were generally conducted using a more lightweight, streamlined approach or a more detailed, comprehensive methodology. Specifically, 11 of these 24 (45.8\%) participants argued that their approach is more detailed, while 50\% stated that they use or have observed the use of a lightweight approach. Those who observed argued that the lightweight approaches were closer to box-ticking exercises (e.g., P05, P09, and P20). Nevertheless, 8 (33.3\%) participants argued that whether a DPIA is lightweight or full-fledged depends on several factors. For instance, P13 stated that it depends \emph{``on a person who is writing it''} while P21 argued that it depends \emph{``on the client and the size of the''} organisation.

\begin{figure}
  \centering
    \includegraphics[width=1.0\columnwidth]{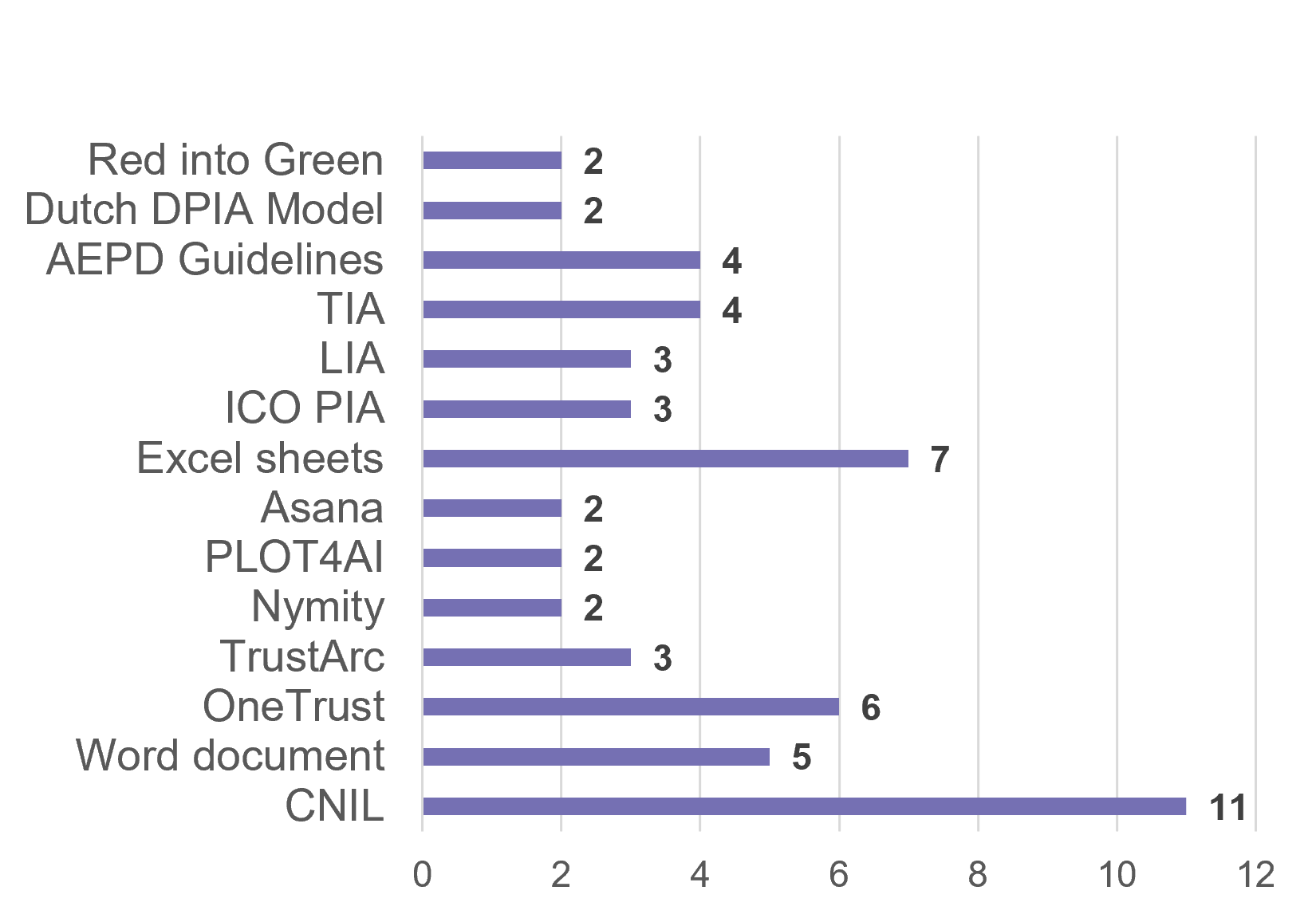}
    \caption{Majority of software and tools mentioned by the participants. LIA stands for Legitimate Interest Assessment, and TIA stands for Transfer Impact Assessment, which may be conducted to support compliance with Article 46 of the GDPR.}\label{fig:tools}
\end{figure}

We also asked participants whether they used software and tools for DPIAs. All participants reported using some form of tool. Figure \ref{fig:tools} depicts the most-mentioned software and tools by participants. As shown, CNIL was the most mentioned, followed by Excel sheets and then OneTrust. In this case, CNIL refers to both the PIA methodology and the PIA software tool. Similarly, the AEPD guidelines were described as a methodology or template. Nevertheless, participants mentioned several tools, ranging from Word documents and Excel spreadsheets to software such as CNIL and OneTrust. 

Importantly, participants identified perceived strengths and shortcomings of CNIL but provided only perceived limitations for OneTrust. Table \ref{tab:cnil} outlines the strengths and weaknesses articulated by some of the participants. The findings suggest that CNIL is perceived as a credible and practically useful methodology, while also revealing some tensions. 

\begin{table}
  \centering
  \caption{Perceived strengths and shortcomings of CNIL}
    \begin{tabular}{p{0.45\linewidth} p{0.45\linewidth}}
    \hline
    \textbf{Strengths} & \textbf{Shortcomings} \\
    \hline \hline
    Useful and established methodology (P17) & \emph{``It's overcomplicated''}(P02) \\ \hline
    The template is more detailed and comprehensive (P13, P17, P20) & Perceived as a box-ticking approach (P05)  \\ \hline
    Useful in practice and \emph{``it's good for people who don't have any template with them or are not so familiar''} (P18) & Found inconvenient when it was translated into Lithuanian (P01)  \\ \hline
    \end{tabular}%
  \label{tab:cnil}%
\end{table}%

The limitations of OneTrust, on the other hand, emerged in several areas. P14 explained that OneTrust and similar tools can be \emph{``horrible''} because they may produce high/medium/low risk ratings that are subjective rather than grounded in objective evidence. In addition, the participant noted that it was expensive, as no one knows how to use it well without spending a lot on consultants. Similarly, P02 argued that they had a problem with OneTrust, in particular the \emph{``internal logic''} and the costs associated with purchasing additional models that require rebuilding internal processes. P11 further stated that OneTrust is \emph{``somehow fostering the idea''} that an organisation needs \emph{``OneTrust professionals''}. This suggests that while the tool is widely known, users are acknowledging some of the limitations emerging in practice. 

Some participants highlighted the general potential benefits of using software and tools. For example, P22 and P03 noted that DPIA tools can help guide assessors through the risk assessment and support report generation, as well as \emph{``help remove that individual experience, individual subjectiveness in the process''}, respectively. However, several participants also identified limitations associated with existing tools. For example, P22 explained that current tools would not generate a proper report, and as such, they would not \emph{``serve the purpose fully''}. Similarly, P06 argued that tools should be sufficiently dynamic to allow assessors to add more questions during assessments, indicating some degree of staticness in the existing tools. In addition, one participant observed, after conducting \emph{``a few proof of concepts and demos with privacy vendors''}, that these tools have \emph{``just a very cute UX and nothing of value''} (P25).

Beyond the tools used to support DPIAs, participants also discussed how they approach the assessment process itself. Specifically, two participants described their approaches to handling a DPIA project. On the one hand, P03 explained that they \emph{``do a reverse engineering [or] bottom-up approach''} in which they could assess, for example, where critical vendors or assets were located. On the other hand, P26 explained that they \emph{``decided to do a top-down approach''}, grouping data types and processes based on information about data use provided by an internal tool. Another participant stated that they conduct umbrella DPIAs. The participant described this as an assessment that allows them to \emph{``enter into negotiations with cloud providers without having a specific organisation as data controller that could be fined by the data protection authority''} (P20). Essentially, this allows them to identify common privacy risks, assess those risks, and provide mitigations. 

Participants also discussed who should be assigned responsibility for the PIAs/DPIAs process. While all participants addressed the question of responsibility, differences emerged. Generally, participants identified business, organisation and process owners, DPOs, legal departments, and privacy and security teams as responsible for the process. Several participants argued that data controllers or project owners should be responsible; for example, P21 explained that \emph{``by law, the data controller is responsible for the DPIA''}, while P01 expressed that \emph{``a data protection officer should not conduct the DPIAs''}, as \emph{``it is a task of [the] data controller''}. Regarding DPOs, as discussed, contrary to P01's statement, several other participants believed that DPOs should be responsible for the DPIAs. However, a few, such as P17, P08, P03, and P13, explained that DPOs are only responsible for, e.g., reviewing and signing off. Nevertheless, one of the participants explained that someone in the business is responsible for the PIA, but because they often lack the necessary knowledge, the DPO needs to support or even lead the work (P06).

In addition, one participant explained how they identify who is responsible. P03 explained that they use a RACI (Responsible, Accountable, Consulted, and Informed) model for this purpose. In particular, they explained that \emph{``RACI is a better approach to identifying who owns it or who is responsible for it versus an ownership question''}. Other participants mentioned that the DPIA's responsibilities should be shared. For example, P08 explained that the responsibility is shared as they \emph{``do not own the project''} other than \emph{``supporting and contributing to it''}. In particular, the participant explained that the project team initiated the process, while the P08 office reviewed the DPIAs, suggesting this shared responsibility. However, while the majority of participants discussed who was responsible, a few others noted that DPIA accountability was undefined. P13, for instance, explained that in a particular case where a DPO was absent, if someone was observed to have an understanding of DPIA, they were given the task to perform, while also involving other team members to support the process.

Another aspect discussed by the participants concerned the timing of DPIAs within a project's lifecycle. Essentially, this was important as several participants observed that it is mostly done as an afterthought (P07, P04, P21, P17, P19). As a result, this would, for example, make things \emph{``more difficult to change''} (P04) or affect the accuracy, as the PIA \emph{``is supposed to document what actually happened''} (P07). As such, participants explained, in their own words, that a DPIA should start early in the DPIA project (P06, P04, P21, P18, P09, P19). As argued by P19, a DPIA should start early in the process, where an organisation has \emph{``an opportunity to consider the problems''}. This means that privacy risks can be captured and addressed from the outset rather than as an afterthought. Apart from starting early, some participants further explained that they conduct a DPIA in an iterative process (P04, P02, P10, P15, P17). P04, for example, described using an Agile approach\footnote{This involves working iteratively, with regular reviews and adjustments based on feedback.} to conducting DPIAs, continuously reviewing and identifying privacy risks throughout each sprint as the project evolves. Hence, PIAs/DPIAs should not be a one-time exercise but a continuous process from the beginning to address risks (P07). 

Finally, participants discussed the extent to which DPIAs are integrated into broader organisational processes, i.e., existing workflows such as risk management and project management. Nine (34.6\%) participants explained that they embed DPIAs into the broader process, while 12 (46.1\%) reported the opposite, as some explained that it was considered a separate exercise, including P02, P03, and P04. However, other participants expressed different views. Two participants stated that embedding DPIA into a broader system was difficult (P04, P20). For example, P20 argued that \emph{``embedding a DPIA in a broader type of managerial overall business impact analysis is usually counterproductive because it [would] reduce the DPIA to 1 or 2 KPIs''}. Another participant noted that whether a DPIA could be integrated into a broader process depended on the organisation, explaining that \emph{``it depends if the company already has a data protection management system in place''}. Nevertheless, several participants (P18, P02, P11, P14) recommended embedding a DPIA into a broader system. For instance, P11 considered this to be the correct approach, while P02 recommended that \emph{``for companies that are growing very fast, it is really important to start as early as possible to create these internal processes in place''}, as this would \emph{``help them to scale up''}.

\paragraph{\textbf{Consulting stakeholders:}} Participants were also asked which stakeholders they engaged in the process and about their relationship with the vendors or third parties when conducting PIAs/DPIAs. Regarding stakeholder engagement, participants discussed, in particular, the engagement of data subjects, external and internal stakeholders, and top management. Of the 26 participants, 21 (80.8\%) discussed the involvement of data subjects, although several differences emerged in their approaches. Eleven participants explained that data subjects were rarely, if ever, involved. For instance, P23 highlighted that they had never involved data subjects because \emph{``the clients didn't see the reason to include''} them, while P03 stated that \emph{``it is hard to get them involved''}. One participant (P08) stated that they had a question regarding the involvement of data subjects, but argued it was not mandatory, at least \emph{``not for every project anyway''}. However, five of the participants explained that they involved data subjects in the DPIA process. P15, for example, mentioned that they had to consult data subjects about their views on healthcare products. Participants also discussed various methods for involving data subjects, including surveys, questionnaires, and focus groups (P04, P21, P25).

Six (23.1\%) participants, however, explained that they engage the employee council as data subjects. In fact, P18 argued that data subjects \emph{``can be external or internal [and] it all depends on the process''} or what a company is providing. This was discussed separately by P03, who further explained that business models would determine the type of data subjects involved, such as B2B and B2C. The participant highlighted that, on the one hand, if their \emph{``client is a B2B client or that's their business model, [one] will not have a direct customer or data subject interaction at all''}. On the other hand, \emph{``when it's a B2C, it's a very retail sort of approach''}. In addition, three (11.5\%) participants explained why they do not involve data subjects. P04 noted that the decision not to involve data subjects sometimes depends on the type of product an organisation is developing and, in some cases, on confidentiality. Similarly, P22 explained that they \emph{``tend not to involve the data subjects because of the potential expenses that may be involved in doing that''}. This is because an industry might have millions of data subjects. Overall, only five (19.2\%) participants recommended involving data subjects in the DPIA process. 

Nineteen (73.1\%) participants discussed their involvement of both external and internal stakeholders, including top management. P21 explained the importance of engaging stakeholders during the DPIA process. The participant stated that one needs \emph{``to very wisely plan their involvement because stakeholders might be valuable for different aspects of the risk assessment''}. This is because they can be knowledgeable about their risk perceptions. However, the size of the organisation was noted to affect stakeholder involvement. P19 argued that stakeholder involvement is complicated, and given the size of an organisation, it would be \emph{``more complicated or less complicated''}. The participant further explained that they were able to \emph{``easily gather all the stakeholders''} in a small organisation, suggesting that doing so in a larger one would be more difficult. However, 6/26 (23.1\%) reported inadequate or a lack of stakeholder and top management involvement. Specifically, P01 noted that \emph{``in 99\% of the cases, just [one] person [is] assigned data protection duties''}. Similarly, P07 argued that they do not usually see multiple people or teams actively working together on a PIA. Instead, one business person is assigned responsibility for completing it. However, the participant stated that occasionally the assigned person might need to contact other teams to obtain specific information, but the process would still be handled primarily by one person. 

On the relationship of third parties in the performance of DPIAs, participants expressed mixed views. Some highlighted the challenges relating to third parties, while others noted that third parties conduct their own DPIAs. Others stated that due diligence should be conducted when dealing with third parties, among other views. In particular, eleven (44\%) participants highlighted that third parties are involved in conducting DPIAs, even if only minimally, as in the case of P24. P06 argued that involving third parties is key to understanding the system and doing the DPIA with due diligence. Another nine (36\%) participants stated that third parties conduct their own assessments. P11 explained that very large vendors \emph{``offer some kind of versions of the DPIA covering their tools because they know they can save a lot of time by doing so, because in those DPIAs there are a lot of elements and information that the clients can use to run their own DPIAs''}. However, P18 argued that while they may contact third parties and ask for DPIAs, they sometimes \emph{``receive really comprehensive impact assessments, but sometimes''} not enough, thus describing them as unacceptable. A few participants (P04, P13, P16, P25) highlighted that organisations should perform due diligence on third parties. P16, for example, explained that their \emph{``procurement department has an internal policy that obliges them to conduct prior tests''} and send questionnaires to \emph{``collect data regarding privacy controls''}. Six (24\%) participants also discussed some challenges related to third parties. P04 explained that a DPIA can become complicated and lengthy due to discussions that may arise between controllers and processors. Furthermore, P14 argued that vendors' PIAs are often intentionally vague, making it difficult for organisations to properly assess vendors' privacy practices and hold them accountable.

\paragraph{\textbf{Risk assessment:}} We also asked participants how they identify and assess risks and controls. All participants responded to this question while expressing different views. Of 26 participants, 22 (84.6\%) described how they identify risks during an assessment. Of the 22 participants, for example, five (22.7\%) discussed identifying risks based on protection goals for privacy and GDPR principles. P18 stated that they \emph{``consider each GDPR principle''}, while assessing whether the processing applies to a particular principle such as data minimisation. Additionally, 7 of 22 (31.8\%) participants took a different approach, identifying potential privacy harms to data subjects. P02, for instance, mentioned that they use Solove's taxonomy of privacy harms to identify risks based on how the clients could harm data subjects. Further, 11 of the 22 (50\%) reported identifying risks through brainstorming, professional experience, and discussions. For example, P04 stated that they use open sessions as they do not often see the whole process, while P09 highlighted that the identification of risks is \emph{``is just largely through individual experience of the process being considered within that organisation''}. Of the 22 participants, five (22.7\%) also discussed the resources they use to identify risks. P20 explained that they \emph{``follow the list of risks identified by the ICO''} while P24 highlighted that they use a compilation of things, including \emph{``some standard risks that the Dutch government lists (...) [and] the ones from FRAIA''}. 

Other participants also expressed different ways of identifying risks. Two participants (P16, P18) explained that they assess risks based on the data category. On the one hand, P16 highlighted that they \emph{``identify risk related to special category of data''}, followed by sensitive data that is not regulated under their law but is presented internally by management and information security departments. The participant explained that they recognise sensitive data as data related to payments, banking details, preferences, and economic behaviour. On the other hand, P18 further stated that in case there is sensitive data, e.g., health data, they assess the risks. In addition, P23 mentioned that their assessment was \emph{``strictly in line''} with the Article 29 Working Party, in which they identified relevant threats and vulnerabilities, assessed the likelihood and impact of different risks, and considered security controls. The participant further explained that they broke down the risks into different types, such as unauthorised access or modification, and data loss, because each risk had a different likelihood and impact, and later combined them \emph{``into one value''}, i.e., risk rating.

Similar to the latter, nine (34.6\%) participants further discussed how they assess likelihood and severity, including risk evaluation criteria. P22 explained that they perform risk scoring quantitatively, by always multiplying the impact by the likelihood \emph{``and all of these are always measured on a scale of 1 to 5''}. Similarly, P11 suggested a quantitative approach calculated using measures from 1 to 5. However, the participant noted that the choice between quantitative and qualitative (i.e., low, medium, and high) depended on the companies. In addition, P04 stated that they use both quantitative and qualitative approaches (i.e., a mixed approach), further explaining that the way risks are identified and scored is \emph{``more''} subjective in practice. However, P05 argued that risk scores do not necessarily provide meaningful precision, with P14 further arguing that it is difficult to decide what to do because privacy harms and benefits affect different people and cannot be compared. As such, unlike the use of either quantitative or qualitative or both, P12 highlighted that they usually assess the severity by asking what would be \emph{``the worst-case scenario and what would happen''}.

Following risk evaluation, 24 (92.3\%) participants explained how they identify mitigations, implement them, and assess whether they have been implemented. Specifically, of these 24 participants, 14 (58.3\%) explained how they identify mitigations and recommend them to clients. P01 and P17 highlighted that they would prepare a list of recommendations. In particular, P17 stated that they would present the list to clients and discuss which recommendations were already in place, which were realistic for the clients to implement, and which they do not have the budget to implement. P25 explained that the organisation had developed a taxonomy of privacy risk informed by the Article 29 Working Party's high-risk criteria and its own experience in assessing projects. The risks identified through this taxonomy were then mapped to controls intended to prevent them. The participant further described an escalation mechanism for high risks that could not be mitigated within the organisation's required timeframe. Such cases were escalated to senior leadership for formal risk acceptance, while risks that could be mitigated on time proceeded through the normal process. In addition, P15 stated that they \emph{``leverage SCF [i.e., Secure Controls Framework] control framework''}, which combines several privacy programs and has its own risk register. However, the participant highlighted that since the framework and register are too broad, they \emph{``created something more custom''}.

Other participants further revealed how they identify mitigations. P10 described that the \emph{``safeguard aspect of a PIA is a huge undertaking and very technical''} and as such, they would bring in privacy engineers or external experts who are good at such issues. Two participants further provided recommendations. On the one hand, P16 stated that controls \emph{``must reflect the decisions made during the assessments''}, arguing that if they did not correspond to the risks, then the work was not well done. On the other hand, P06 recommended that organisations should \emph{``not limit the mitigation actions only to this specific processing activity or assessment''} and \emph{``should result in some training, specific tailored trainings for those who are involved in data management or the process management''}.

Participants also described how these mitigations are implemented, including who is responsible for implementation, and how their implementation is verified. A few participants (P15, P23, P26) described their implementation process. P15 argued that while different teams would be working on a DPIA and the implementation of measures, certain risks would emerge that are easy to mitigate and address, for example, by updating a privacy notice to increase transparency. However, the participant also mentioned that they would \emph{``be part of the implementation''}, particularly for one-off projects that require help taking a product all the way from development through to market launch. In addition, P23 described how they implemented mitigations through risk treatment plans. The risk treatment plan involved accepting, avoiding, reducing, or sharing the risks. Where risk reduction was selected, the participants explained that the next step was to determine specific controls to address the risk. Furthermore, P26 explained that the choice of measures to implement was influenced by factors such as cost and the potential impacts on the entire IT ecosystem, and as such, the organisation sometimes adopted measures that were considered sufficiently effective and feasible, such as restricting certain uses of data, providing free training, or \emph{``proactively deleting data''} from selected environments. The participant also noted that projects could proceed while mitigations were still being implemented, with the organisation accepting the remaining risk for a period of time and documenting the ongoing activities.

Furthermore, of the 24 participants, 11 (45.8\%) explained who is responsible for the implementation. The participants shared different views, but one thing that emerged from most of them is that someone was assigned to implement the measures. For example, P08 noted that they \emph{``usually assign the risk to someone''}, even though they would be responsible for certain things like updating a privacy notice, while P25 explained that they assign \emph{``a person responsible for the process and their engineering team''}. Similarly, P11 discussed presenting a list of actions to a function (e.g., an IT department or privacy team) and, importantly, holding them accountable. These views were also shared by P22, who highlighted that they \emph{``map the risks to the relevant persons (...) either the IT team or the legal team, or DPO''}, and \emph{``ensure that they [are] (...) implementing the mitigation measures''} in a proactive manner. However, P06 argued that DPOs are \emph{``not responsible for implementing the mitigation actions''}. In addition, eleven (45.8\%) participants explained how they verify that the mitigations have been implemented. Different views emerged; for example, some stated that they followed up (P03, P08, P09, P12, P18, P19), had limited visibility to check implementations (P02), had no visibility (P05), and/or were not involved in the implementation (P13, P23). P20 described an approach whereby institutional negotiating leverage was used to secure audit rights, enabling independent auditors to draft questions to assess compliance, for example, whether agreed data retention periods were being followed, and whether inappropriate access to data had occurred, including measures that had been implemented.

Finally, one participant (P17) raised a broader concern regarding the duplication of risk assessment. P17 suggested that organisations could conduct a single risk assessment covering all processing activities, rather than conducting separate assessments that may overlap. In particular, the participant noted that organisations already conduct a risk assessment under Article 24 and that an additional DPIA under Article 35 could be burdensome, particularly for SMEs. Instead, the participant proposed a single risk assessment that provides an overall picture of the risks and controls that could be applied to address them. The participant further suggested that data protection authorities could provide a common methodology for conducting such assessments, referring to CNIL's approach as an example. 


\paragraph{\textbf{Documentation and review:}} Here, all participants discussed PIA/DPIA documentation and dissemination, while 23 of the 26 participants also discussed the revision and updating of PIAs/DPIAs. Under documentation and dissemination, participants discussed several aspects, including finalising the DPIA process, assessing whether a DPIA is ``good enough'', and dissemination of DPIA reports. Five of the 26 (19.2\%) participants discussed how a DPIA process is finalised. Specifically, participants explained that once they have completed a DPIA, they deliver it to either a DPO or the person responsible for approving it.

Regarding how participants assess whether a DPIA is ``good enough'' after completing the assessment, and whether they have quality criteria for making this determination, participants offered different views. The largest group of participants (12/26, 46.1\%) reported that they do not have an explicit quality criterion for assessing a good DPIA. P07, for instance, argued that \emph{``it is very hard to know how accurate the responses are in the PIA without testing''}, adding that many privacy teams lack people with auditing expertise. Similarly, P21 explained that a good DPIA template can make sure that all the required information is entered, but cannot guarantee that the information people enter is correct, for example, the legal basis for data processing. As such, checking that requires expertise and experience. 

In addition, other participants shared how they assess whether a DPIA is good enough. Nine of the 26 (34.6\%) described that good DPIAs are reflected in risk identification and mitigation. P02 argued that when someone understands the processing activity, identifies the risks, and finds mitigation measures for them, this reflects a criterion for a good DPIA. Another nine (34.6\%) participants explained that they decided a DPIA is good based on its completeness, e.g., \emph{``if it's answering all your questions''} (P06). Several other participants explained how they determined this. P11 and P24 argued that a DPIA would be good if it met the requirements under the GDPR, specifically Article 35. In addition, three other participants (P08, P09, P19) stated that the criteria would be based on the DPIA's clarity, understandability, and auditability. For instance, P09 explained that if \emph{``a logical and auditable thought trail is available for independent analysis''}, then the DPIA could be considered good enough. 

Participants also discussed the dissemination and disclosure of DPIA reports. In particular, the discussion focused on whether participants disseminate DPIA reports internally and/or externally, what information they communicate to data subjects, and who within the organisation reads the reports. Regarding external dissemination, 21 out of 26 (80.7\%) reported mixed views. Specifically, ten of the 21 (47.6\%) participants argued that DPIAs are not published externally. P25, for instance, stated that they have never published a DPIA publicly. Similarly, P17 highlighted that they recommend publishing a summary, as they have not seen a complete DPIA report published. While this was the case, three out of 21 (14.3\%) discussed the importance of disseminating DPIA reports externally. P02 argued that it is \emph{``not obligatory to publish a DPIA''}; however, they advised their clients to do so, as {``it could be very beneficial for them to publish''} to promote transparency toward customers. Two participants expressed similar views, noting that external publication could demonstrate transparency (P06, P15). Nevertheless, one participant explained that they do not publish DPIA reports externally, but \emph{``only upon request because these are confidential information internal to the organisation''} (P03). In addition, 2 participants (P09, P24) described factors that could result in limited or no external dissemination. In particular, P09 argued that DPIA is essentially an internal document, and organisations may publish one once a project is mature, but public disclosure is generally not considered a standard requirement. Furthermore, P24 explained that they \emph{``do not share the DPIAs externally most of the time unless the DPIA is drafted in a joint controller relationship''}. However, five of 21 (23.8\%) reported publishing DPIA reports externally. P08 and P18, for example, stated that they publish them externally on their websites. 

Participants were asked whether they internally disseminated the DPIA reports. Fifteen of the 26 (57.7\%) participants responded; only one (P02) of them stated that they do not share internally. P02 highlighted that once the DPIA is completed, it is not shared internally; instead, it is set aside and ultimately forgotten. The remaining participants shared their views regarding whom they share the report with internally. For instance, P08 stated that they \emph{``share the copies of the DPIA [with] the project team for awareness''} and \emph{``encourage them to look at the risk sections and make sure that they update or (...) reduce the risk for that project''}. Similarly, P06 pointed out that they share the DPIA report with stakeholders \emph{``as part of implementation of the mitigation actions''}. P03 further explained that they disseminate to \emph{``the right stakeholders to create visibility''}, such as to the \emph{``CTOs, CIOs, risk officers, CROs, the C-level, as well as (...) maybe at some operational levels, depending on where the need is''}. 

In addition, we asked participants (20/26, 76.9\%) which information from DPIA reports they considered important to present to the data subjects. Table \ref{tab:info_datasubjects} summarises their considerations. As noted, participants considered it important to communicate the personal data being collected and processed, as well as the risks and mitigations. Participants also described methods that would be used to communicate to data subjects, such as through a privacy policy or notice, or through blogs, where they would use \emph{``simple language [to] summarise the findings''} (P20). However, one participant (P26) argued that providing too much information could give attackers leverage. In addition, P14 also stated an opinion that data subjects \emph{``do not want to be bothered''}, and instead, they would  \emph{``want to work with a company that stewards their data and treats it appropriately within their expectations for whatever service''}.

\begin{table}
  \centering
  \caption{Information from PIA/DPIA reports to be communicated to data subjects.}
    \begin{tabular}{p{0.25\columnwidth} p{0.65\columnwidth}}
    \hline
    \textbf{P.ID} & \textbf{Information} \\
    \hline \hline
    P01, P03, P07, P11, P13, P26 & Through privacy policies or notices, e.g., data being collected (P03)  \\ \hline
    P02, P05, P15, P18, P21, P24 & Data processing activities, e.g., communicating about processing activity, data storage, data retention, data subject rights (P24) \\ \hline
    P04, P06, P09, P11, P17, P22 & Risks and mitigations overview, e.g., identified and mitigated risks relevant to data subjects before explicit consent (P17) \\ \hline
    P13 &  Information from \emph{``official request to DPO''} \\ \hline
    P20, P25 & Writing blogs to summarise and share information, e.g., planning to \emph{``give information on data subject rights''} (P25) \\ \hline
    \end{tabular}%
  \label{tab:info_datasubjects}%
\end{table}%

Furthermore, participants discussed whether PIAs are periodically reviewed. Participants shared different views on this, including the review timeframe (i.e., the time required to revise a DPIA) and the treatment of the DPIA as a living document. Focusing on the latter, six of the 23 (26.1\%) participants stated that a DPIA should be kept as a living document to allow for revision. For example, P02 stated that they always advise that DPIAs should not be treated as a one-time document to be completed and forgotten, but should be kept up to date and reviewed regularly. As such, a few participants (P04, P07, P20) explained that DPIAs get revised or refreshed. In particular, P04 highlighted that the DPIAs get revised, \emph{``especially for more sensitive applications''}. However, the participant stated that \emph{``not all organisations do that''}. Nine of the 23 (39.1\%) participants explained triggers for new revisions. P19, for instance, stated that \emph{``if something changed because, for example, they have a new vendor, they needed to immediately communicate that and assess again that vendor and the new risk, if any''}. In addition, P02, P13, P15, P18, P22, and P23 argued that a significant change in the processing activity or the introduction of new technology warranted a revision. Furthermore, twelve of the 23 participants (52.2\%) discussed the DPIA review timeframe. Table \ref{tab:review_timeframe} outlines participants' articulation of the review timeframe.

\begin{table}
  \centering
  \caption{PIA/DPIA review timeframe as articulated by participants.}
    \begin{tabular}{p{0.25\columnwidth} p{0.65\columnwidth}}
    \hline
    \textbf{P.ID} & \textbf{Review timeframe} \\
    \hline \hline
    P01 & 3 years \\ \hline
    P02 & Mostly 1 year (although rare); but should be revisited 4 or 5 months \\ \hline
    P04 & 1 or 2 years \\ \hline
    P08, P11, P21, P22, P24, P25 & 1 year \\ \hline
    P16 & Quarterly reviews \\ \hline
    P17 & Less than three years \\ \hline
    P19 & Every 6 months \\ \hline
    \end{tabular}%
  \label{tab:review_timeframe}%
\end{table}%

As shown in Table \ref{tab:review_timeframe}, most participants stated that a DPIA should be reviewed annually. However, P02 argued that, while it's reviewed once a year most of the time, this is not frequent enough. As such, the participants recommended that the DPIAs should be reviewed every 4 or 5 months. P02 further explained that in practice it is difficult to get clients to do this because they believe they have completed the DPIA and therefore tend to forget about it. Following this, several participants (P01, P09, P14, P21, P23, P26) highlighted that DPIAs are not reviewed after completion, while P11 and P17 argued that revisions are dependent on the organisation. In addition, P10 explained that revising a PIA is \emph{``not strictly a legal obligation''}. 

Two participants (P14, P16) also discussed sanity checks. On the one hand, P14 argued that people doing ``sanity checks'' would be needed to review PIAs and assess whether the information provided makes sense, while more formal audits should be conducted periodically, particularly for high-risk PIAs, to verify that the information in the PIA reflects the actual processing activities. On the other hand, P16 argued that external auditors should be hired to verify the quality of the data privacy office's work, assess its processes, and check their controls, and \emph{``measure up efficiency of assessments''}. 




\subsubsection*{Sub-theme T2.3: Concerning Trade-off Assessment}
We further asked the participants how they address trade-offs between data protection goals and/or other values, e.g., data minimisation vs transparency. Participants (25/26, 96.1\%) reported different views, from conducting an actual trade-off assessment to trade-off assessments being implied as difficult or impossible. Nine of the 25 (36\%) stated that they assessed trade-offs. For instance, P04 argued that they do trade-off analysis even though they had not heard it done in practice, especially among professionals in AI. The participants explained that it would be easier to include a trade-off analysis as part of a matrix in the process and to hold a dedicated session with different stakeholders to work through the trade-offs. In addition, P08 highlighted that they address trade-offs, especially between data minimisation and AI model training (e.g., model accuracy, fairness, or robustness). While they noted that they strongly encourage data minimisation, they balanced the trade-off through necessity and proportionality. However, if they realised the whole project was high risk, they would decide to stop it. 

A further seven of the 25 (28\%) explained that the decision regarding how to assess trade-offs should be taken by the company. P06 noted that when trade-offs arise, they can be escalated to senior managers, who need to support the organisation's privacy culture. Similarly, P02 highlighted that as an advisor, it is \emph{``up to the company to decide what they will do (...) or how they will resolve the situation''}. Another five of the 25 (20\%) described how such trade-offs should be managed. In particular, P03, P18, and P24 discussed this in relation to AI and data minimisation. P24 argued that one should not have to use the minimal amount of data; instead, one should use the amount of data necessary and proportionate to achieve the given goal. P03 highlighted the need for a business to determine whether to use synthetic data or anonymised datasets, or to train models using personal data. The participant argued that, depending on the choices made by an AI-based company, it can evaluate risks to data subjects and other privacy risks when conducting a PIA. Lastly, P18 explained that there is indeed a conflict between the need for vast amounts of data to make the AI models work well and data minimisation. However, the participant noted that the massive data collection has already happened to develop the foundational models, and that cannot be undone. As such, the participant mentioned that organisations should focus on reducing privacy risks in AI systems and, in particular, conduct training and raise employee awareness of how to use the tools.

One participant (P25) noted that it is difficult to assess trade-offs, while P01 argued that they have never assessed trade-offs and that attempting to do so would only increase the burden on the DPIA. Another participant (P12) explained that they \emph{``do not like these trade-offs very much''}; instead, they block the activity if the risks identified are too high.

\paragraph{\textbf{Engagement with supervisory authority:}} 
The engagement with the supervisory authority was discussed by very few participants (P09, P18, P22). The general discussion was that, in the event of high or residual high risk, the controller is required to consult with the regulator. For instance, P18 explained that if a high risk is not mitigated, a data protection authority needs to be consulted for advice. Similarly, P09 argued that when \emph{``there's a residual high risk identified after mitigations have been applied, then there's an obligation on the controller to submit that DPIA''} to the regulator for advice. 



\hfill \break 
\noindent\fbox{
    \begin{minipage}{.95\linewidth}
        \setlength\parskip{1em}
        \textit{\textbf{Key Takeaways:}\\
        $\rightarrow$ DPIA practice was noted to lack a standard template/methodology, as the majority of participants developed bespoke methodologies. \\
        $\rightarrow$ Data subjects were rarely involved or considered as part of stakeholders; however, a few participants reported consulting them.  \\
        $\rightarrow$ Despite differing views, several participants noted that considerations of necessity and proportionality were important, particularly when weighing trade-offs and ensuring data minimisation. Yet, most do not employ any method for such analysis.
        } 
    \end{minipage}
}

\subsubsection{T3. Perceived Benefits and Challenges of PIAs/DPIAs}\label{benefits and challenges}
\begin{figure}
  \centering
    \includegraphics[width=0.85\columnwidth]{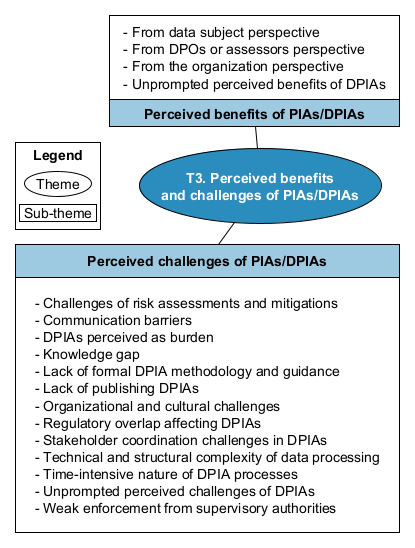}
    \caption{Overview of theme 3 and its sub-themes.}\label{fig:theme_3}
\end{figure}

Figure \ref{fig:theme_3} illustrates the theme and sub-themes, including summarised codes for each sub-theme, which partly address RQ3.

\paragraph{\textbf{Peceived benefits:}} All participants discussed perceived benefits of DPIAs. Most predominantly, participants discussed the benefits of conducting a DPIA. As stated in Section \ref{findings}, the responses were extracted from direct questions, while others emerged organically during discussion of other topics. Of the 26 participants, 24 (92.3\%) discussed the benefits of conducting a DPIA mostly from an organisation's perspective. Several participants (11/24, 45.8\%) described compliance and governance as the main benefits. For example, P24 stated that a DPIA \emph{``allows you to ensure privacy by default, by design, compliance with laws and regulations, and internal policies''}. Several others (8/24, 33.3\%) noted that it provides organisational visibility. P14, P15, and P19 argued that conducting a DPIA provides visibility into processing activities and their associated potential risks. P14, for instance, explained that the main benefit is that it helps management discover what is actually happening inside a large organisation, ensuring that they \emph{``have visibility into activities and in particular potentially risky activities''}. P15 further explained that it gives an overview of the \emph{``risk environment''}. 

In addition, six of the 24 (25\%) participants explained that the benefit of conducting a DPIA is creating data protection awareness. P08, P24, and P25 discussed that it increases awareness within the organisation, especially among stakeholders. Furthermore, five of 23 explained that DPIA benefits extend beyond compliance. For example, P22 stated the organisation would gain public trust, as well as give it a competitive advantage, as it would be \emph{``meeting thresholds prescribed by law''} and as such, it is \emph{``likely to have a more predictable legal regulatory compliance''}. In addition, P02 argued that a DPIA \emph{``can be a good relationship builder (...) especially if the companies will (...) publish their DPIA''}. P06 supported this by highlighting that, when conducted correctly, the DPIA can be published externally to enhance transparency and serve as a communication tool for external stakeholders. Furthermore, a few participants (P04, P15, P16) explained that conducting a DPIA can benefit an organisation by preventing organisational harm. P16 argued that it \emph{``helps to avoid fines and penalties''}. However, the participant explained that carrying out a PIA does not eliminate the possibility of being fined; it \emph{``mostly reduces the risk''}.

Beyond perceived benefits toward the organisations, 15 of the 26 (57.7\%) participants noted benefits toward data subjects. Nine participants explained that conducting a DPIA enables risk identification and implementation of mitigations. P10 argued that the essence of conducting a PIA is to \emph{``analyse a program or activity, identify privacy risks, and then mitigate and try to reduce those risks''}. As such, P09 stated that \emph{``the idea and application of appropriate risk analysis can only benefit a particular project if due regard is given to potential obstacles, problems, and, more importantly, mitigations and answers to those problems''}. In addition, P07 argued that conducting a DPIA early in the process helps identify potential problems before deployment. This enables efficient resource allocation and prevents problems from occurring in the first place. 

Furthermore, seven of the 15 (46.7\%) participants discussed DPIA benefits related to the protection of data subjects. Specifically, P17, P21, and P22 argued that conducting a DPIA should protect the fundamental rights of the data subjects. P22, for instance, explained that a benefit for data subjects should be the protection of their human rights. In addition, P19 stated that one of the benefits is to \emph{``think what can happen and what the risks for''} the data subjects are and how they can be minimised. 


\paragraph{\textbf{Perceived challenges:}} Similar to benefits, all participants expressed and discussed perceived challenges. However, the extent of the challenges discussed was much greater than the benefits, and as such, we tabulated this as outlined in Table \ref{tab:perceived_challenges}. The responses were extracted from direct questions, while others emerged organically during the discussion of other topics. While some of the perceived challenges are self-explanatory, others contain nuances that warrant further discussion. 

\begin{table}
  \centering
  \caption{Perceived challenges of PIAs/DPIAs. Note: P(\%) stands for participants' percentages.}
    \begin{tabular}{p{0.8\columnwidth}c}
    \hline
    \textbf{Challenge} & \textbf{P(\%)} \\
    \hline \hline
    Organisation and cultural challenges & 88.5\% \\ \hline
    Challenges with stakeholder coordination and communication & 53.8\% \\ \hline
    Challenges of risk assessments and mitigations & 50\% \\ \hline
    DPIAs perceived as a burden & 34.6\% \\ \hline
    Perceived weak enforcement by data protection authorities & 26.9\% \\ \hline
    Lack of formal DPIA methodology and guidance & 26.9\% \\ \hline
    Knowledge gap & 19.2\% \\ \hline
    Time-intensive nature of DPIA processes & 19.2\% \\ \hline
    DPIAs as compliance formalities/documentation exercise & 15.4\% \\ \hline
    Challenge to define the right scope & 15.4\% \\ \hline
    Technical and structural complexity of data processing & 15.4\% \\ \hline
    Late initiation of DPIA & 11.5\% \\ \hline
    DPIAs not revisited or updated & 7.8\% \\ \hline 
    Lack of publishing DPIAs & 7.8\% \\ \hline
    Perceived low value of DPIAs in practice & 7.8\% \\ \hline
    Complex to gather a lot of information for DPIAs & 3.8\% \\ \hline
    \end{tabular}%
  \label{tab:perceived_challenges}%
\end{table}%

As outlined in Table \ref{tab:perceived_challenges}, a majority of participants perceived organisational and cultural challenges. The major challenges highlighted are: (i) Assessments focusing on risks to the organisation rather than data subjects, (ii) limited resources affecting the assessment, (iii) limited resources hindering revisions and updates, (iv) lack of privacy knowledge among personnel, (v) organisations failing to understand the value of a DPIA, and (vi) organisational resistance to initiating PIAs. 

Participants (50\%) often described challenges concerning risk assessment and mitigation, including implementation. The main point of discussion was that organisations were not treating DPIAs as effective risk management processes. Instead, participants highlighted difficulties in identifying and assessing privacy risks, and ensuring that identified risks and proposed mitigation measures are actually implemented. The latter was noted as some participants stated that responsibility, stakeholders' dynamics (i.e., employees prioritise core job responsibilities), and follow-up are often lacking (P01, P04, P07, P21, P24). In addition, participants argued that risk assessment can be subjective (e.g., P03, P14, P16), does not translate into actual consideration of the risk (P14), and requires input from stakeholders beyond legal or privacy teams (P04). Furthermore, P05 emphasised that privacy risks extend beyond cybersecurity and should be considered from the data subject's perspective, highlighting this as a challenge, as many stakeholders were only concerned about external hackers.

Regarding perceived limited enforcement by data protection, participants reported their experiences with supervisory authorities. The major challenge discussed was the role of enforcers, as some participants observed a lack of follow-up from supervisory authorities on the DPIAs conducted (P07, P08, P17, P20, P21). P06 shared their experience, arguing that the supervisory authority does not do the work well, adding that they always complain \emph{``about lack of resources''}. Given this, P20 suggested improvements to the prior consultation process. In particular, the participant argued that the process should be improved by requiring supervisory authorities to respond to questions within a mandatory, reasonable timeframe. The reason is that organisations do not need to endure long delays caused by staff shortages from the supervisory authority. Furthermore, P21 argued that regulators should not treat organisations as if their only purpose is to comply with privacy law. Instead, they should understand legitimate business needs, be \emph{``much open to discussions''}, and provide clear and practical instructions, and \emph{``not just fine companies randomly''}.

Participants also discussed a challenge related to knowledge gaps in their teams. In particular, this challenge was observed based on who was/is conducting the PIA/DPIA. P04 highlighted that, in certain cases, \emph{``the one performing a DPIA is just a lawyer without technical knowledge''}, and as such, it becomes difficult to discuss technical details. This lack of technical details was noted by P10, who argued that while they might be knowledgeable in technical details, there are cases in which they lack certain knowledge. Some examples of knowledge gaps reported by participants were: not knowing how to conduct an assessment (P08), not understanding the value of a risk assessment (P09), and certain stakeholders not knowing about the project being assessed (P19). Regarding the latter, P19 explained that a mismatch can emerge between stakeholders' knowledge of the project and what is being assessed in the DPIA. The participant attributed the mismatch to a lack of knowledge, noting that some of the individuals involved may not be sufficiently familiar with the project. 

Scoping and threshold assessment challenges were also highlighted by some participants (P03, P11, P14, P18). In particular, they explained that organisations have difficulty scoping DPIAs. Specifically, P03 argued that organisations find it challenging to determine at what level of granularity to conduct the DPIA, e.g., whether to conduct it from the bottom up or at an abstract level. P14 further explained how the challenge emerged. The participant argued that technological systems often support multiple business processes (e.g., an email system \emph{``connected to 50 different processes''}, thus making it difficult to define clear boundaries of a DPIA. Participants further emphasised that conducting DPIAs is complicated by complexity, opacity, and the constantly changing nature of data processing (P11, P14, P19, P26). P11, for example, explained that complex contexts, such as ad tech and data monetisation, could be difficult to understand \emph{``who is interacting with whom, where the data is''}. This could also include app developers who integrated SDKs, as well as other factors that introduce opacity (P14).

\hfill \break 
\noindent\fbox{
    \begin{minipage}{.95\linewidth}
        \setlength\parskip{1em}
        \textit{\textbf{Key Takeaways:}\\
        $\rightarrow$ DPIAs were viewed as focusing on organisational risks rather than risks to data subjects. \\
        $\rightarrow$ Participants perceived weak and resource-constrained enforcement by data protection supervisory authorities. \\
        $\rightarrow$ DPIAs were viewed as tools that could provide visibility into an organisation's processing activity and gain insight into the risk environment. 
        } 
    \end{minipage}
}
\subsubsection{T4. Interplay with AI Regulation}

\begin{figure}
  \centering
    \includegraphics[width=1.0\columnwidth]{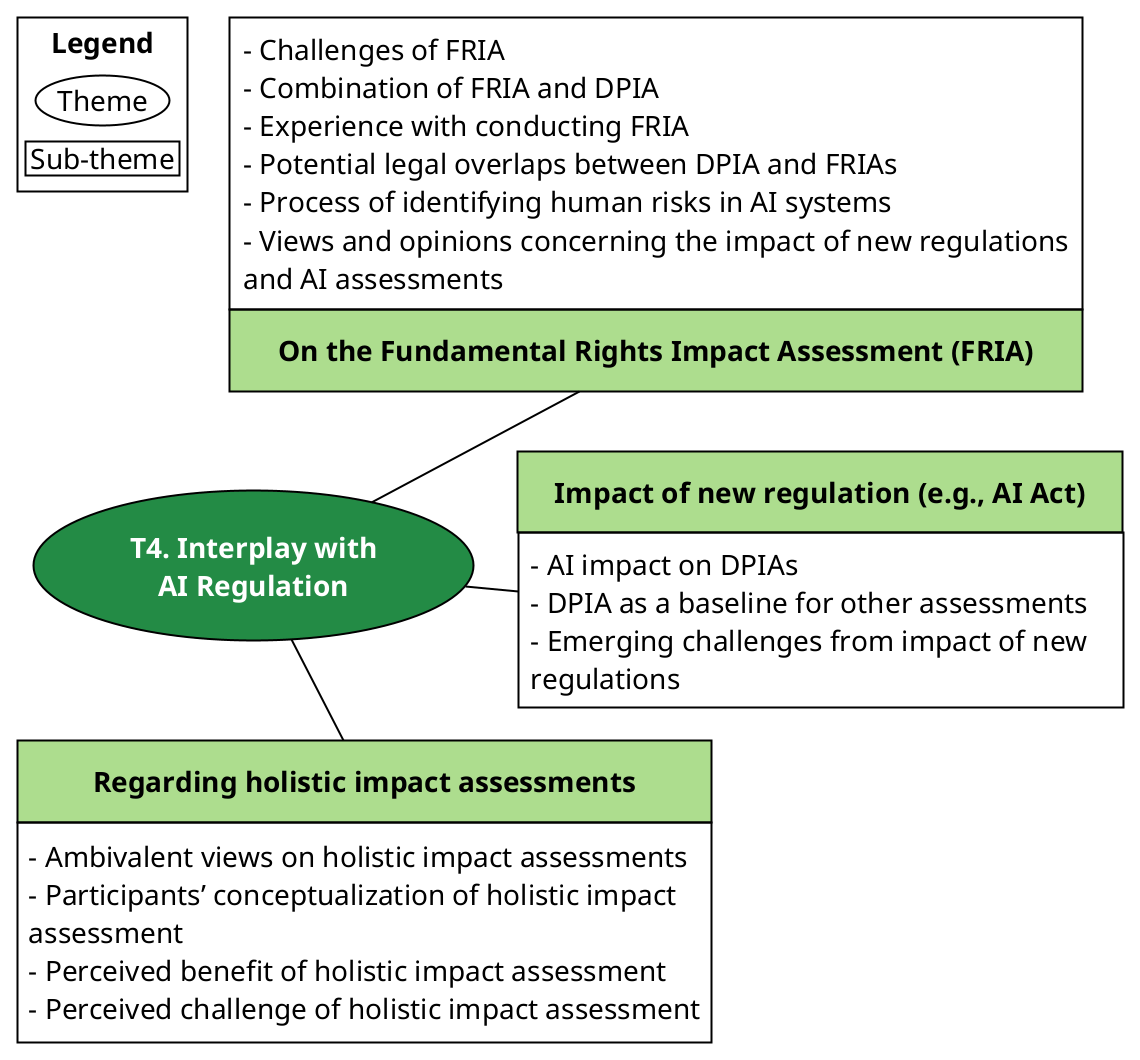}
    \caption{Overview of theme 4 and its sub-themes.}\label{fig:theme_4}
\end{figure}

Figure \ref{fig:theme_4} illustrates the theme and sub-themes, including summarised codes for each sub-theme, which partly address RQ2.

\subsubsection*{Sub-theme T4.1: Impact of New Regulations (e.g., AI Act)}
We asked participants about their views on whether new AI regulations (e.g., the AI Act) and the introduction of other impact assessments could affect the future of PIA/DPIA routines. In total, 24 (92.3\%) discussed this aspect, with several views and opinions emerging. Of the 24 participants, 20 (83.3\%) discussed the potential impact of the AI Act on DPIAs, though some noted that the regulation would have no impact. Specifically, 16/20 (80\%) of the participants suggested that the EU AI Act and other emerging regulations, such as \emph{``the general EU Digital Strategy [that] has lots of new legislation, like the Cyber Resilience Act, NIS2, the Data Act''} (P18), could influence how DPIAs are conducted. P17 argued that the EU AI Act is likely to influence DPIAs; however, they stated that they are already considering many of the AI issues in DPIAs. P09 explained that the new regulations may introduce additional requirements, but the DPIA process should remain risk- and principle-based, noting that established principles, the identification of specific risks, and the implementation of appropriate mitigations could reduce risks associated with emerging technology.  A further point of interplay with the EU AI Act highlighted how new regulatory requirements shape assessment practices, with DPIAs discussed as a baseline for other assessments, including those related to DORA and FRIA (P03, P05, P21). P21, for instance, stated that DPIAs already have a broader connection to FRIAs, as FRIAs can be built on previously conducted DPIAs that focus on high-risk AI systems.

Additionally, several participants (P03, P04, P05, P08, P11, P19) expressed concerns about the impact of new AI regulations on PIA/DPIA practices. In particular, discussions about the difficulty of conducting additional impact assessments under the AI Act were raised. Specifically, participants mentioned problems including \emph{``assessment fatigue''} (P03), the complexity and resource intensity of conducting multiple assessments (P04, P19), additional paperwork (P08), and overlapping legislation (P11). These challenges were argued to result from the growing number of regulatory assessment requirements, including FRIAs, alongside existing assessments such as PIAs/DPIAs. In addition, P05 explained a challenge regarding \emph{``what's actually going on and how decisions are made''} in AI systems, noting that it would be extremely difficult to conduct impact assessments.

Nevertheless, a few participants (P01, P06, P12, P20, P24) expressed that the AI Act has no impact on how DPIAs are conducted, because the EU GDPR already requires organisations to assess high risks associated with the processing of personal data and new technologies. As such, P01 and P12 saw no additional impact from the AI Act, particularly where the system does not trigger an existing GDPR requirement for a DPIA. Similarly, P20 argued that the DPIA remains highly relevant because AI systems frequently involve personal data and decisions about individuals, whereas the AI Act's impact assessment requirement applies only to a more limited category of AI systems. However, P24 stated that, while there is no impact, \emph{``people are becoming more aware that just because you are legally allowed to do something doesn't make it ethical, and that there might be a difference there, and that you should also take ethics into account''}. P06 took a more positive view of additional assessment requirements. The participant argued that more assessments are beneficial because people working in IT \emph{``are extremely reluctant to do [the] right things''}. As such, additional requirements can protect \emph{``end users''}.

\subsubsection*{Sub-theme T4.2: On Fundamental Rights Impact Assessments (FRIAs)}
\begin{figure}
  \centering
    \includegraphics[width=1.0\columnwidth]{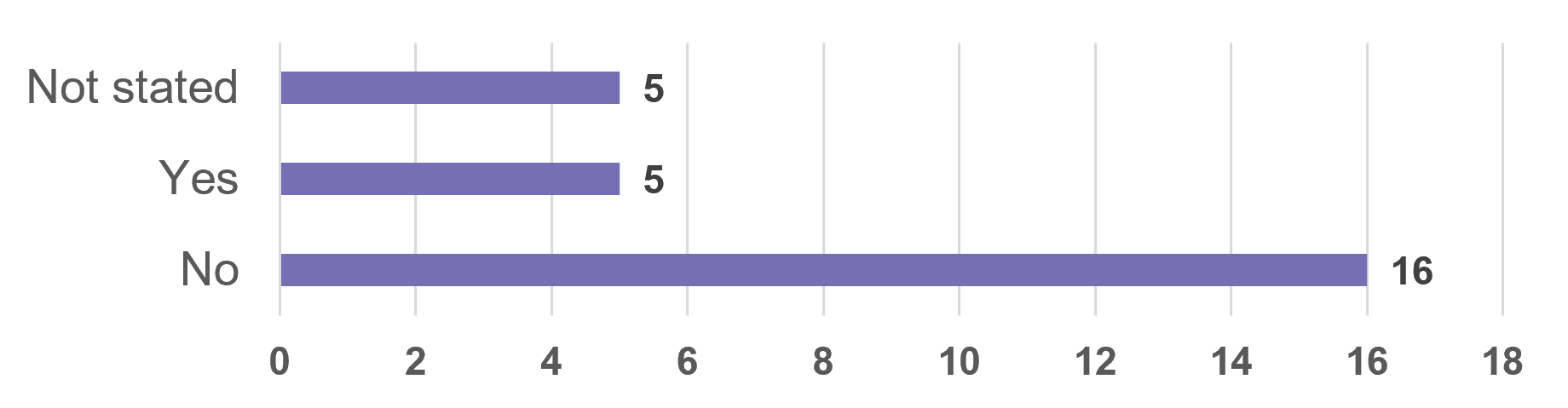}
    \caption{Participants response on conducting FRIA.}\label{fig:fria}
\end{figure}

We asked participants whether they were familiar with Fundamental Rights Impact Assessments (FRIAs), whether they had any experience with the assessment, and whether they were being discussed or used within their organisation. Figure \ref{fig:fria} illustrates the number of participants who reported their experience with conducting FRIAs, who at the time of the interview had not conducted FRIAs, and who did not state whether they conducted FRIAs. As noted, 16 of the 26 (61.5\%) participants reported being unfamiliar with, or having not conducted, FRIAs at the time of the interview. P17 stated that they have not conducted FRIAs because the AI models they worked with did not qualify as high-risk AI systems. P25 also mentioned that they had not conducted one but had only \emph{``gave a workshop''} regarding FRIA, while P15 highlighted that, at the time of the interview, it was not a legal requirement. 

In addition, while 5 out of 26 (19.2\%) participants did not state whether they had conducted or had experience with FRIAs, another 19.2\% (i.e., P03, P04, P15, P18, P20) mentioned that they had indeed conducted FRIAs. Specifically, P04 and P20 explained that they conducted an FRIA using the Impact Assessment on Human Rights and Algorithms (IAMA) and the Fundamental Rights and Algorithms Impact Assessment (FRAIA), respectively, whereas P03, P15, and P18 did not specify any methodology. However, it is important to note that although P15 conducted an FRIA in cooperation with a European-based organisation, they explained that conducting FRIAs was not a legal requirement in their jurisdiction at the time of the interview. 

Furthermore, several other participants (16/26, 61.5\%) expressed differing views regarding FRIAs. In particular, six of the 16 (37.5\%) participants identified challenges associated with FRIA. Specifically, P03, P12, and P20 highlighted the lack of established guidance and methodologies, with P03 noting that \emph{``there is not enough guidance at the moment, from EDPB or ECJ, because there are not many cases that have come in this space yet''}. Similarly, P20 observed that \emph{``there is no accepted methodology''} and thus \emph{``waiting for the AI office and the national authorities''} to provide guidelines. Furthermore, P03 noted concerns about the burden emerging from multiple assessments, which they referred to as assessment fatigue, describing FRIAs as \emph{``just one more thing to do''}. P21 noted the \emph{``power imbalance''} between providers and deployers, explaining that although deployers are responsible for conducting the FRIA, they may have limited ability to address risks identified in AI systems developed by more powerful providers. In addition, P25 highlighted the \emph{``overlap with other legal areas''} and noted that assessing fundamental rights could amount to \emph{``invading some other team's scope''}, suggesting that responsibilities for fundamental rights assessments may overlap across different teams.

Despite these challenges, eight of the 16 (50\%) participants explained the combination of DPIA and FRIAs, further highlighting the impact of new regulations. One participant explained that they \emph{``rather use that something that exists in an organisation to build upon that or the area instead of like, recreating the whole thing''} (P03). The combination of DPIA and FRIA was also motivated by the need to save resources (P01, P11), with P11 arguing that if an organisation enlarges \emph{``the scope of the DPIA to also capture the FRIA requirements''}, they would \emph{``save money and time''}. Nevertheless, P11 further argued that DPIAs and FRIAs should not be assumed to be identical assessments, but one should recognise the applicable AI Act requirements within the DPIA process where relevant, while acknowledging that the two frameworks have different regulatory focuses. This means that one focuses on high-risk processing of personal data, while the other \emph{``applies to AI systems and models''}. 

Regarding the identification of risks to fundamental rights, P05 emphasised the value of involving diverse stakeholders with different perspectives to identify and discuss risks, arguing that the same approach used to assess data protection risks could be applied to other fundamental rights. P26 described how their organisation used red-team and blue-team approaches to test AI systems for potential risks to individual fundamental rights and to develop appropriate mitigations. In addition, P07 argued for privacy engineering in AI systems rather than just an AI risk assessment, emphasising that risks and potential harms should be considered throughout the entire development cycle. 

However, the difficulty of measuring human rights risks emerged as a particular concern. P05 argued that measuring such risks \emph{``is certainly even more difficult than for data protection''}. 

\subsubsection*{Sub-theme T4.2: Regarding Holistic Impact Assessments}
We also asked about the idea of having more Holistic Impact Assessments (HIAs), where several assessments, for example, privacy, security, human rights, etc., could be combined into one. Of the 26 participants, 23 (88.5\%) discussed this while expressing different views. Four participants (P04, P06, P07, P23) discussed their conceptualisation of holistic impact assessments. P04 described the conceptualisation of HIA as based on a threat modelling library that helps a person build Trustworthy AI while providing \emph{``a holistic overview of what''} could go wrong. In particular, the participant illustrated this by describing the use of multiple categories covering security, safety, data governance, human rights, human oversight, accessibility, and privacy while emphasising the need to adapt the assessment to the specific use case and focus on the most relevant risks. Similarly, P06 described it as part of the approach they apply. Essentially, the participant highlighted the importance of considering contextual factors affecting individuals, such as \emph{``family situation [and] place of living''}, rather than limiting risk assessment to requirements explicitly stated in the EU GDPR, further explaining that one \emph{``should have a holistic view of those things''}. Furthermore, P23 explained how they combined the EU GDPR and DORA within a single assessment process, recognising areas of overlap while maintaining distinctions in how consequences are assessed. P07 also explained the integration of different approaches and tools, including privacy engineering and threat modelling, while also questioning the assumption that an assessment \emph{``is a one-time exercise''}. 

Another four participants (P03, P08, P16, P25) expressed different opinions while viewing HIAs as a promising approach. However, they emphasised that such an approach should make an assessment dynamic, risk-based, and focused on relevant areas. P03 argued that, while they agree with HIAs, the approach would not work on Excel sheets, as people would not answer a thousand questions. Instead, the participant explained that there should be a tool that would be used to \emph{``consolidate all the different assessments and questions, and make it dynamic in a way that you are only evaluating certain areas of assessment''}, when only a criterion is met. P08, in addition, explained that HIA would be great if all phases or sections are \emph{``all fully accounted for''}; however, the participant argued that that would \emph{``make the document maybe 100 pages''}. P16 further emphasised that the purpose should be \emph{``to address the risks''} and not necessarily about a \emph{``concrete framework''}. One participant (P25) expressed uncertainty about the approach's practicality in a large organisation with specialised teams assessing the products. In particular, the participant argued that while they get their \emph{``risk partners in the same room to carry out a holistic analysis of a project sometimes, day-to-day-wise, it doesn't work like that''}.

Despite different views, several participants stated that the HIA approach has benefits and challenges. In particular, eight of 23 (34.8\%) and thirteen of 23 (56.5\%) participants discussed benefits and challenges, respectively.  On the one hand, participants expressed benefits, for instance, being able to \emph{``have knowledge about everything that's going on''} in an organisation (P02), understanding not only which risks are connected, but also how one risk may amplify another, thus making risk analysis easier and meaningful (P04), and avoiding information overlaps by conducting a single assessment rather than several (P14, P15, P17).  On the other hand, participants described challenges such as potential difficulty in defining what a holistic assessment is or how it should be conducted (P01), risk of HIAs becoming \emph{``too high level to be properly useful as a development tool for any given area of potential responsibility''} (P09), risk of making \emph{``the procedure kind of longer''}, and difficulty combining assessments into one, as there are many regulations such as the GDPR, DORA, AI Act, Data Governance Act, and NIS (P23).


\hfill \break 
\noindent\fbox{
    \begin{minipage}{.95\linewidth}
        \setlength\parskip{1em}
        \textit{\textbf{Key Takeaways:}\\
        $\rightarrow$ The impact of new regulations, especially the EU AI Act, has been suggested by combining FRIA requirements in the DPIA process, including building FRIAs on top of DPIAs.  \\
        $\rightarrow$ The emergence of several assessments, such as FRIAs, has been perceived to potentially cause assessment fatigue. \\
        $\rightarrow$ HIAs show some potential, as a few participants reported applying such an approach, while others discussed advantages, albeit with challenges.
        } 
    \end{minipage}
}

\section{Discussion}\label{disc}





Although the GDPR has transformed DPIAs from a largely voluntary privacy practice into a formal regulatory requirement, our interviews with 26 experienced professionals across 17 countries show that formalisation has not produced a uniform practice. Instead, professionals continue to negotiate the meaning, scope, ownership, methodology, timing, and value of impact assessments within organisational contexts. The findings suggest that the quality of a PIA/DPIA depends less on the existence of a particular template or tool than on whether the assessment is substantively integrated into organisational decision-making.
The emergence of AI-related fundamental-rights assessments further intensifies this challenge, creating a need for integrated but modular impact-assessment practices rather than simply more compliance documentation.

Furthermore, the study presents additional findings on the state of the practice of PIAs and DPIAs in a post-GDPR landscape. The GDPR has not only intensified efforts in the EU, essentially through mandatory assessments for high-risk systems, but also cast its net globally due to its extraterritorial scope. Several participants outside Europe point to the GDPR and equivalent requirements for PIAs in their own jurisdictions, e.g., the US, Brazil, and Canada. To some extent, this is needed because non-EU entities must comply with the GDPR if they offer goods or services to individuals in the EU or monitor their behaviour, regardless of the organisation's physical location. However, many professionals also stated that the EU GDPR has been seen as a gold standard in privacy regulation, and its principles and requirements have affected other countries' legal frameworks, including requirements for PIAs or DPIAs.

Although much can be discussed regarding the study's findings, in the following subsections, we focus on central topics that we recognise as practical challenges to overcome. These challenges around PIAs/DPIAs concern the timing and decision-making, the standardisation of organisational practices, organisational ownership and stakeholder engagement, and the potential for holistic impact assessments, as discussed below.

\subsection{Timing and Decision-Making}
Participants' accounts also highlight the importance of timing. The European Commission emphasises that a DPIA should be conducted before processing begins and treated as a living tool rather than a one-off exercise~\cite{europeancommission_dpia}. Our findings similarly suggest that assessments undertaken after key decisions have been made may have less opportunity to influence the underlying project and may instead function primarily as documentation of decisions already taken. This reinforces the broader understanding of impact assessment as an \textit{ex ante} and iterative governance mechanism. The finding also suggests that the existence of a completed DPIA is an imperfect indicator of its practical effectiveness: its influence depends partly on whether the assessment is embedded in decision-making at a stage when its findings can still affect outcomes~\citep{wright2012state}.

\subsection{Formalisation Without Standardisation}
A central finding is that the formalisation of DPIAs under the GDPR has not led to standardised organisational practice. Participants described substantial variation in methodologies, thresholds, tools, responsibilities, and ways of conducting assessments. This is consistent with \cite{iwaya2024privacy}, who found heterogeneous PIA practices and limited empirical evidence regarding their practical effectiveness. Similarly, the work of \cite{hansen2025illuminating} found that ad-hoc approaches and manually produced DPIA reports remain common despite the GDPR's formal requirements. Our findings extend this work by suggesting that this heterogeneity is not simply a consequence of organisations failing to adopt established methods. Professionals actively adapt, combine, or develop approaches in response to organisational context. Recent regulatory developments involved the introduction of a DPIA template by the EDPB~\citep{EDPB2026}, illustrating efforts to promote harmonisation in regulatory practice. However, a common template may not necessarily eliminate differences in how professionals interpret risks, exercise judgement, and operationalise assessment requirements within their organisational context. Nevertheless, the DPIA template provides a starting point for harmonisation. 


Interestingly, we found that many professionals are familiar with technical standards and tools that can support both the conduct and management of PIAs and DPIAs. Still, such technical standards and tools are not widely implemented in the organisations' internal risk management process. Although participants have criticised some software tools for PIAs and DPIAs, there is still room to improve and leverage these tools to achieve better standardisation of organisational practices.

\subsection{Organisational Ownership and Stakeholder Involvement}
The findings further indicate that the distribution of responsibility and expertise within organisations shapes PIA and DPIA practice. Participants described different arrangements involving privacy professionals, legal teams, security specialists, project teams, and management. This supports previous observations that DPIAs are organisational processes rather than activities performed solely by privacy specialists. For instance, \cite{hansen2025illuminating} highlighted the importance of multi-stakeholder involvement and found that such involvement is often insufficiently reflected in practice. Our findings similarly suggest that clear ownership and appropriate stakeholder involvement are important conditions for moving PIAs/DPIAs beyond documentation towards substantive decision support.

Particularly when reflecting on organisational ownership, it is useful to consider that PIAs and DPIAs, as one specific type of privacy practice, can be strongly influenced by the company's organisational privacy climate, i.e., the shared perception of how behaviour related to privacy is rewarded, supported, and expected within an organization~\citep{hadar2018privacy}. Leadership sets the tone for privacy values, and effective cultures require privacy training, clear roles, and transparency to move beyond reactive compliance~\citep{iwaya2022organisational}. This means that top management has an important role in promoting and encouraging privacy practices, including PIAs/DPIAs, rather than expecting isolated professionals or external consultants to champion such practices within the organisation. As argued by~\citep{wright2012findings}, PIAs do need champions, but also high-level support and an embedded privacy culture.

\subsection{Integrating or Conducting HIAs}
It is worth noting that the ideas of holistic or integrative impact assessments have already been a topic of discussion for many years~\citep{wright2013integrating, morrison-saunders2014strengthening, mantelero2018ai}.
In this study, we also presented professionals with the idea of holistic assessments as an approach to more effectively utilising resources and addressing overlapping assessment work.
Professionals noted that identifying the right people to conduct a multifaceted assessment is likely the principal challenge, which is consistent with previous findings in the literature~\citep{wright2013integrating}.
A person or team well-equipped to assess multiple dimensions, such as privacy, security, and human rights, is rare.
Furthermore, a holistic assessment also inherently increases the complexity and demands more resources (time, money, and stakeholders).
This would be especially burdensome for SMEs that already run in a resource-constrained organisational structure.
The same burden goes to supervisory authorities, which would be tasked with auditing and reviewing such holistic assessments.
For such assessments to be realistic, they must be tactfully structured so that the benefits of avoiding redundant work are accompanied by sufficient depth of analysis across multiple facets compared with performing assessments in isolation.
Supervisory authorities also need to confirm that such assessments would be sufficient to demonstrate compliance.

\section{Threats to Validity}\label{threats-to-validity}
In this section, we discuss the main threats to validity in the study and outline its limitations. We use the classification of threats to validity from \cite{runeson2009guidelines} to analyse internal and external validity, construct validity, and reliability, as discussed below.

First, we consider the study's \emph{internal validity}, which focuses on examining causal relationships, particularly whether unexamined factors may influence the results~\citep{runeson2009guidelines}. Since interviews rely on self-reporting rather than direct real-life observation, participants may be constrained by memory limitations, brief timeframes, and a natural tendency to overemphasise specific topics relative to their everyday reality. To mitigate this bias, we encouraged participants to ground their responses in specific real-life cases and used frequent follow-up questions to assess the depth and accuracy of their reported practices.

\emph{External validity} refers to the extent to which findings can be generalised~\citep{runeson2009guidelines}. We do not make any claims about the generalizability of our findings; a sample of 26 participants across various organisations does not provide a statistically representative cross-section of privacy professionals. Furthermore, while the sample reflects diverse professional backgrounds and includes perspectives from outside Europe and North America, it remains predominantly Western. Therefore, future studies should explore the views of professionals from underrepresented regions around the world.

\emph{Construct validity} pertains to the alignment between the operational measures (i.e., the instruments used) and the research questions being investigated~\citep{runeson2009guidelines}. The primary instruments utilised in this study, including the screening survey and interview guide, are provided in Appendix~\ref{interview-guide} of the supplementary materials. To enhance these tools, two external researchers and a Data Protection Authority (DPA) privacy commissioner reviewed the interview guide before we piloted it in an initial interview. After making minor adjustments, all authors finalised and approved the instruments.

Lastly, \emph{reliability} concerns whether the data and analysis are dependent on the specific researchers involved~\citep{runeson2009guidelines}. Although the research team oversaw the project collectively from the outset, the first author led the thematic analysis. To enhance analytical rigour, an external research fellow who specialises in qualitative security and privacy research collaborated on the initial stages of analysis. All three researchers used an open-coding strategy and subsequently synthesised their preliminary codebooks. Moreover, the entire author team, drawing on interdisciplinary expertise in security, privacy, software engineering, and qualitative methodologies, engaged in iterative reviews to reach consensus on code generation, thematic development, and the overarching framework.

\section{Conclusion}\label{conclusion}
A substantial amount of research has been conducted on PIAs and DPIAs~\citep{iwaya2024privacy}, and systematic reviews identify numerous methodologies that have been rigorously evaluated~\citep{wairimu2024slr}. However, there appears to be a misalignment between research and practice, as privacy professionals often do not utilise consolidated PIA methods. This disconnect is partly due to the tension between the need for comprehensive, rigorous assessments and the reluctance of organisations and clients to allocate the necessary resources, viewing them as extra costs rather than investments, and overlooking the benefits. In this regard, it is encouraging that regulations such as the GDPR make DPIAs mandatory for high-risk systems, as many organisations perceive them as mere obstacles to overcome.

Furthermore, this study also shows that although GDPR has formalised DPIAs, their implementation still varies considerably between organisations. Effective PIAs and DPIAs depend on templates and tools; however, early integration into decision-making, clear responsibility, and meaningful stakeholder involvement are necessary. Overall, the findings highlight the need for more consistent yet flexible approaches that make impact assessments practical and useful for managing privacy risks.

Notwithstanding this, future work is needed to better understand views in other regions of the globe, particularly among under-researched populations in the Global South. More research is also needed to evaluate existing methods for FRIAs (and HRIAs), as more organisations will be expected to conduct them, at least in the EU under the AI Act. The integration of impact assessments also remains a challenge in striking a balance among sufficient depth across various facets of assessment, increased stakeholder engagement, and the fulfilment of legal obligations. Authorities also need to provide more guidance on whether such holistic or integrative assessments would be considered valid for demonstrating compliance.

\subsection*{Funding}
Samuel Wairimu and Leonardo H. Iwaya have been funded by the Swedish Knowledge Foundation (KK-stiftelsen) via the H\"{o}g program, financing the Operationalizing Human Rights in the Design of AI-supported Digital Health (OPEHRA) research project. Leonardo H. Iwaya is also partly funded by Region V\"{a}rmland (Grant: RUN/230445) and the European Regional Development Fund (ERDF) (Grant: 20365177) in connection with the DHINO 2 project, and by Vinnova (Grant: 2018-03025) via the DigitalWell Arena project. Ala Sarah Alaqra received no funding support for this research.

\subsection*{Acknlowedgements}
The authors would like to thank Simone Fischer-Hübner, Marit Hansen, and Meiko Jensen for early feedback on the interview guide. We would also like to thank all the professionals who took time out of their busy schedules to participate in our interviews.




\clearpage 



\appendix

\section{Appendix A -- Interview Guide}\label{interview-guide}
The following is the interview guide used in this study. Qs are the questions asked, and the bullet points are potential follow-up questions to probe further responses. Note that questions from existing related work (Table \ref{tab:related-work}) have been referenced and integrated into our interview guide.

\subsection{Introduction}
Welcome and thank the participant. Introduce our aim, confirm whether the consent form has been submitted, and highlight that the responses are voluntary. The emphasis was on professionals' experiences rather than on the organisation's specific inner workings to avoid any hesitancy about confidentiality breaches. 

\subsubsection{RQ1: General Perspectives}
\begin{itemize}
    \item Q1. Could you tell us about your professional role, years of experience and education?
    \begin{itemize}
        \item Do you have any privacy-related certification?
        \item Have you received any training in privacy-related (internal/external)?
        \item Which area of practice? Which organisation, type, or sector?
        \item Number of years of experience in the privacy field? We may ask for other demographics, including age group and gender, etc.
        \item How did you end up in this position? \citep{parks2011understanding}
    \end{itemize}
    \item Q2. Tell us about your experience of PIA in practice.
    \begin{itemize}
        \item What is your experience with PIAs and DPIAs?
        \item Have you received any training more specifically for PIAs or DPIAs?
        \item What approach do you know/use for conducting PIAs?
        \item How much experience do you have with performing PIAs? \citep{vandercruysse2021dpia}
        \item Since when has your organisation conducted PIAs? \citep{van2017privacy}
        \item How many PIAs are conducted in your organisation? \citep{van2017privacy}
    \end{itemize}
    \item Q3. What is a PIA or DPIA, according to you?
    \begin{itemize}
        \item Is it similar to information security? Why? Why not? \citep{parks2011understanding}
        \item Do you see differences between PIAs and DPIAs?
    \end{itemize}
    \item Q4. What are the main privacy regulations in your work or for your organisation? What are the main privacy standards?
\end{itemize}

\subsubsection{RQ2: How PIAs are Conducted}
\begin{itemize}
    \item Q1. How do you usually go about conducting (planning and executing) a PIA for a project or organisation?
    \begin{itemize}
        \item How do you start a PIA? Is there an organisational process for it?
        \item In which phase or phases in the product and/or information system development is the PIA conducted? \citep{van2017privacy}
        \item In which cases does your organisation conduct / not conduct a PIA (is there a threshold)? \citep{van2017privacy}
        \item How is it decided whether or not a data processing activity will be subject to a PIA? \citep{vandercruysse2021dpia}
        \item Is there a guideline on how to conduct a PIA? On which methodology or standard is it based? \citep{van2017privacy}
    \end{itemize}
    \item Q2. Has the PIA been built into the project management of another business process? \citep{van2017privacy}
    \item Q3. Who is responsible for the PIA process?
    \begin{itemize}
        \item Who conducts the PIA (an individual or a team, which functions are represented)? \citep{van2017privacy}
        \item Who else do you believe should be responsible for the PIAs?
    \end{itemize}
    \item Q4. Who are the stakeholders involved in the PIA process? Are there any key persons in the organisations or departments?
    \begin{itemize}
        \item Who else is involved in the PIA? Do you consult any stakeholders? Who are the stakeholders?
        \item Do you involve data subjects or their representatives?
        \item Are there people who can help you perform PIA? Such as internal or external consultants, in-house experts, or the DPAs?
        \item Do you involve lawyers or other technical privacy and security experts?
    \end{itemize}
    \item Q5. How is the relationship with your vendors or third parties in the performance of PIAs (e.g., data processors, joint data controllers)? Do they participate? Do they have their PIAs for their solutions?
    \item Q6. In your opinion, what type of PIA methodology is usually employed, a more lightweight/streamlined version or a more detailed/full-fledged PIA?
    \item Q7. Do you use any software system or platform for managing and conducting PIAs?
    \item Q8. How do you identify and assess risks and controls? (threat analysis)
    \begin{itemize}
        \item Do you use any types of catalogues or knowledge bases?
        \item What is the usual terminology? Do you refer to threats, risks, controls, impacts, harms, etc.?
        \item (If they use detailed terminology)
        \begin{itemize}
            \item How do you assess privacy risk? How do you assess likelihood, severity, and impact? What influences the level of privacy impact? \citep{ferra2020challenges}
            \item How does the sensitivity of data influence the impact? How do you assess the sensitivity of data? \citep{ferra2020challenges}
            \item How do user expectations influence the impact? How do you assess user expectations? \citep{ferra2020challenges}
            \item How does the harm caused influence the impact? How do you assess the harm caused by privacy violations? \citep{ferra2020challenges}
            \item What other factors influence the impact of privacy risks? \citep{ferra2020challenges}
        \item What gaps do you see in privacy risk assessment? \citep{ferra2020challenges}
        \end{itemize}        
    \end{itemize}
    \item Q9. How do you address conflict between data protection goals and/or other values? E.g., data minimisation vs transparency and fairness, data minimisation vs safety or energy-saving/sustainability.
    \item Q10. How do you know that a PIA is good enough? Do you have quality criteria for a PIA?
    \begin{itemize}
        \item Is the quality of the PIA assessed? By whom? \citep{van2017privacy}
        \item Is somebody assigned to manage the PIAs (e.g. the privacy officer) \citep{van2017privacy}
    \end{itemize}
    \item Q11. How do you go about implementing the measures and controls that you identified? (Implementation)
    \begin{itemize}
        \item Do you have an auditing process for PIAs? What is it?
        \item How and when is it monitored if the mitigating measures of PIA are implemented during the development phases? \citep{van2017privacy}
        \item In your opinion, who should be responsible for implementing the measures identified through a PIA?
    \end{itemize}
    \item Q12. Are PIAs periodically revised (is this an obligation)? \citep{van2017privacy}
    \item Q13. Did the outcome of the PIA result in changes in the (specifications of the) information system? \citep{van2017privacy}
\end{itemize}

\subsubsection{RQ3: Benefits, Challenges and Public PIAs}
\begin{itemize}
    \item Q1. What are the biggest benefits of conducting a PIA, according to you?
    \begin{itemize}
        \item Do you consider informing the public/end-users a factor?
        \item Do you consider competitive edge as a factor?
    \end{itemize}
    \item Q2. What are the biggest challenges for conducting a PIA, according to you?
    \begin{itemize}
        \item Do you consider limited resources as a factor?
        \item Do you consider regulations as a factor?
    \end{itemize}
    \item Q3. How do you disseminate the PIA reports? Are they publicly available?
    \item Q4. Who do you think will read a PIA conducted by your organisation?
    \begin{itemize}
        \item Has it been presented to and impacted relevant decision-makers/management in the organisation?
    \end{itemize}
    \item Q5. Which information from the PIA do you consider important to be presented to the data subjects?
    \item Q6. Are you familiar with public registries for PIAs? What do you think of them? Do you believe that you or your organisation would be willing to take part in this?
    \item Q7. How do you see the new regulations (e.g., AI Act) for AI/ML systems and further demands for impact assessments? Do you think these may impact PIA routines in the future?
    \item Q8. Are you familiar with Fundamental Human Rights Impact Assessments? If so, what is your experience with them, and are they being discussed or used in the organisation? What about holistic impact assessments?
\end{itemize}

\subsubsection{Comments and closing remarks}
\begin{itemize}
    \item Q1. Is there anything else you would like to share?
    \item Q2. Why did you agree to participate?
    \item Q3. Would you like us to share the results of the study with you?
    \item Thank the participants. Provide the contact information for follow-up.
\end{itemize}

\printcredits

\bibliographystyle{cas-model2-names}

\bibliography{cas-refs}



\end{document}